\documentclass[twocolumn]{jpsj3}
\usepackage{graphicx}
\usepackage{color}
\usepackage{bm}
\title{%
Phase Diagram of a Non-Hermitian Chern Insulator:
Destabilization of Chiral Edge States and Bulk--Boundary Correspondence
}

\author{%
Yositake Takane
}

\inst{%
Graduate School of Advanced Science and Engineering,\\
Hiroshima University, Higashihiroshima, Hiroshima 739-8530, Japan
}

\recdate{ \hspace{50mm} }

\abst{%
A non-Hermitian Chern insulator with gain/loss-type non-Hermiticity
shows a peculiar gap closing; when a nontrivial Chern insulator phase
changes to a gapless phase, conduction and valence bands are
combined into one band owing to destabilization of chiral edge states.
A previous recipe of non-Hermitian bulk--boundary correspondence is
insufficient for this system since such a gap closing is beyond its scope.
Here, we revise the recipe by taking the peculiar gap closing into account
and apply it to the non-Hermitian Chern insulator.
We demonstrate that the bulk--boundary correspondence holds in the system
including a destabilization of chiral edge states.
A phase diagram derived from the bulk--boundary correspondence
is shown to be consistent with spectra of the system.
}

\begin{document}
%%\sloppy
\maketitle

\section{Introduction}

Bulk--boundary correspondence is a characteristic feature that manifests
itself in various topological systems.~\cite{thouless,kohmoto,kane,fu1,
moore,roy,ryu1,fu2,hasan,qi}
A topological number defined in bulk geometry
under a periodic boundary condition (pbc) plays a central role
in the original scenario of bulk--boundary correspondence proposed
by Hatsugai~\cite{hatsugai} and Ryu and Hatsugai.~\cite{ryu2}
The topological number governs the presence or absence of
topological boundary states in the boundary geometry
under an open boundary condition (obc);
if it is nontrivial (trivial), topological boundary states appear (disappear)
in the boundary geometry.
Note that the original scenario employs the bulk and boundary geometries.

Inspired by an extension of quantum mechanics
to the non-Hermitian regime,~\cite{hatano,bender,brody}
extensive studies have been conducted on
non-Hermitian topological systems.~\cite{hu,esaki,malzard,
yuce1,t_lee,leykam,xu,ashida,xiong,gong,shen,kunst1,yao1,yao2,zyuzin,
lieu,alvarez,yoshida1,kawabata1,yin,kawabata2,longhi1,yokomizo1,okuma1,
song,herviou,okugawa1,yoshida2,c_lee,papaj,kunst2,imura1,zhou1,
kawabata3,torres,yoshida3,borgnia,yang,yokomizo2,zhou2,e_lee,wu1,wu2,
mochizuki,koch,yuce2,mostafavi,wang,zhao,he,yokomizo3,imura2,kondo,
kawasaki,yoshida4,takane1,yoshida5,yokomizo4,okugawa2,xie,zhu,bessho}
Bulk--boundary correspondence is an important subject of these studies.
It was demonstrated that a straightforward application of
the original scenario fails to describe the topological nature
of some non-Hermitian topological systems.~\cite{t_lee,xiong}

A reason for the breaking of the bulk--boundary correspondence is
a non-Hermitian skin effect;~\cite{yao1,okuma2,zhang,yi,longhi2,
kawabata4,li,okuma3,longhi3,guo}
eigenfunctions in a non-Hermitian system under the obc
are localized near a boundary of the system.
This effect is hidden in the system under the pbc.
Indeed, all eigenfunctions under the pbc extend over the entire system.
Therefore, a topological number defined in the bulk geometry under the pbc
never reflects the non-Hermitian skin effect in the boundary geometry.
Several scenarios without using the bulk geometry
have been proposed.~\cite{yao1,yokomizo1,kunst1,borgnia}
Although these scenarios can avoid difficulty concerning
a non-Hermitian skin effect,
they do not share the notable feature of the original scenario
that a topological number defined in closed geometry
governs the presence or absence of topological boundary states.

To restore the bulk--boundary correspondence
in non-Hermitian topological systems,
another scenario has been proposed in Refs.~\citen{imura1} and \citen{imura2}.
This scenario employs boundary geometry under the obc
and bulk geometry under a modified periodic boundary condition (mpbc).
Let $|\Psi^{R}\rangle = \sum_{n=1}^{N}\varphi(n)|n\rangle$ be
a right eigenstate of a one-dimensional lattice system consisting of $N$ sites.
The mpbc is defined by $\varphi(N+n) = b^{N}\varphi(n)$,
where $b$ is a positive real constant.~\cite{imura1}
The wavefunction $\varphi(n) = \beta^{n}$ satisfies this boundary condition
when $\beta = b e^{ika}$, where $a$ is the lattice constant.
The case of $b = 1$ corresponds to the ordinary pbc.
If $b > 1$ ($b < 1$), the mpbc selects wavefunctions
that exponentially increase (decrease) with increasing $n$.
This behavior of $\varphi(n)$ resembles
that caused by a non-Hermitian skin effect.
Therefore, by using the bulk geometry under the mpbc, we can take
the non-Hermitian skin effect into account
in calculating a topological number.

The scenario in Refs.~\citen{imura1} and \citen{imura2} succeeds
in describing the bulk--boundary correspondence
in one-dimensional non-Hermitian topological insulators.
The present author~\cite{takane1} gave a recipe for executing the scenario
in two-dimensional systems and applied it to
a Chern insulator with gain/loss-type non-Hermiticity,~\cite{yao2}
obtaining a phase diagram in the boundary geometry.
The phase diagram separates three regions:
a topologically nontrivial Chern insulator phase,
a topologically trivial insulator phase, and a gapless phase.
In Ref.~\citen{takane1}, the phase boundary between
the nontrivial and trivial regions and that between trivial and gapless regions
are shown to be consistent with spectra in the boundary geometry.
However, the phase boundary between the nontrivial and gapless regions
is not fully consistent with spectra.

In this paper, we examine the bulk--boundary correspondence
in the Chern insulator with gain/loss-type non-Hermiticity
by using a revised recipe.
A key observation is that the inconsistency reported in Ref.~\citen{takane1} is
caused by a peculiar gap closing in the boundary geometry:
if chiral edge states linking conduction and valence bands are destabilized
by non-Hermiticity, the two bands are combined into one band
without an obvious gap closing.
This is beyond the scope of the recipe in Ref.~\citen{takane1}.
We thus revise the recipe of the bulk--boundary correspondence such that
it can take the peculiar gap closing into account
and show that the revised recipe enables us to describe
the topological features of the system without any inconsistency.
Indeed, we obtain a phase diagram that is fully consistent with
spectra in the boundary geometry.

In the next section,
we present a tight-binding Hamiltonian for the non-Hermitian Chern insulator
with gain/loss-type non-Hermiticity~\cite{yao2}
and specify two symmetries of it.
Complex spectra of this system are constrained by these symmetries.
In Sect.~3, we introduce right and left eigenstates of the Hamiltonian
as a preparation for theoretical analyses.
In Sect.~4, we consider chiral edge states in the geometry with
the obc in one direction and the mpbc in the other direction.
This consideration reveals important characteristics of chiral edge states
in the boundary geometry, which are used in the following two sections.
Section~5 is the central part of this paper.
We give a revised recipe for executing
the scenario of the bulk--boundary correspondence
and obtain a phase diagram in the boundary geometry by using it.
In Sect.~6, we justify the resulting phase diagram by comparing it
with spectra of the system in the boundary geometry.
The last section is devoted to a summary and discussion.

\section{Model and Symmetries}

We adopt a tight-binding model for non-Hermitian Chern insulators
on a square lattice with lattice constant $a$.
The Hamiltonian is given by
$H = H_{\rm d}+H_{x}+H_{y}$ with~\cite{yao2}
\begin{align}
   H_{\rm d}
 & = \sum_{m,n} |m,n \rangle
                \left[  i\gamma\left(\sigma_{x}+\sigma_{y}\right)
                      + M\sigma_{z}
                \right]
                \langle m,n| ,
         \\
   H_{x}
 & = \sum_{m,n} |m+1,n \rangle
                \left[ \frac{i}{2}v\sigma_{x} - \frac{1}{2}\sigma_{z}
                \right]
                \langle m,n|
              + {\rm h.c.} ,
        \\
   H_{y}
 & = \sum_{m,n} |m,n+1 \rangle
                  \left[ \frac{i}{2}v\sigma_{y} - \frac{1}{2}\sigma_{z}
                  \right]
                \langle m,n|
              + {\rm h.c.} ,
\end{align}
where $M$, $\gamma$, and $v$ are real parameters,
$m$ and $n$ specify the location of each site
in the $x$- and $y$-directions,
and $\sigma_{x}$, $\sigma_{y}$, and $\sigma_{z}$ are the $x$-, $y$-,
and $z$-components of Pauli matrices, respectively.
$|m,n \rangle$ and $\langle m,n|$ consist of spin-up and -down components:
\begin{align}
  |m,n \rangle
  & = \left[ \begin{array}{cc}
               |m,n \rangle_{\uparrow} & |m,n \rangle_{\downarrow}
             \end{array}
      \right] ,
     \\
  \langle m,n|
  & = \left[ \begin{array}{c}
               {}_{\uparrow}\langle m,n| \\
               {}_{\downarrow}\langle m,n|
             \end{array}
      \right] ,
\end{align}
where $|m,n \rangle_{\sigma}$ and ${}_{\sigma}\langle m,n|$
with $\sigma = \uparrow$, $\downarrow$
are, respectively. column and row vectors.
If the number of sites in the system is $N_{\rm site}$,
the column and row vectors consist of $2N_{\rm site}$ components,
where the factor $2$ reflects the spin degrees of freedom.
In $|m,n \rangle_{\sigma}$ and ${}_{\sigma}\langle m,n|$,
only one component corresponding to the $(m,n)$th site with $\sigma$ is $1$
and other components are $0$.
They satisfy ${}_{\sigma}\langle m,n| = {}^{t}|m,n \rangle_{\sigma}$ and
\begin{align}
   {}_{\sigma}\langle m,n|m',n' \rangle_{\sigma'}
    = \delta_{m,m'}\delta_{n,n'}\delta_{\sigma,\sigma'} .
\end{align}
As a consequence, $H$ is a $2N_{\rm site} \times 2N_{\rm site}$ matrix.
The Hamiltonian $H$ is normalized such that
the coefficients of $\sigma_{z}$ in $H_{x}$ and $H_{y}$
become $-\frac{1}{2}$ and therefore $H$ is dimensionless.
The non-Hermiticity of the system is characterized by $\gamma$.
For simplicity, we focus on the case of
$0 \le M$ and $0 \le \gamma$ with $v = 1$ throughout this paper.
The model becomes topologically nontrivial when $M < 2$
in the Hermitian limit of $\gamma = 0$.

In the remainder of this section, we point out two symmetries that
$H$ possesses.
The first symmetry is
\begin{align}
  \sigma_{x}H^{*}\sigma_{x} = - H ,
\end{align}
where $\sigma_{x}$ acts on $2 \times 2$ spin space.
From this symmetry, we can show that, if $|\Psi^{R}\rangle$ satisfies
$H|\Psi^{R}\rangle = E |\Psi^{R}\rangle$,
then $\sigma_{x}|\Psi^{R}\rangle^{*}$ satisfies
$H\sigma_{x}|\Psi^{R}\rangle^{*} = -E^{*} \sigma_{x}|\Psi^{R}\rangle^{*}$.
That is, an eigenstate of $H$ with an eigenvalue of $E$ is paired with
its counterpart with the eigenvalue of $-E^{*}$
if ${\rm Re} \{ E \} \neq 0$.
The second symmetry is
\begin{align}
  {}^{t}(UP)({}^{t}H) UP = -H ,
\end{align}
where $U = i\sigma_{y}$ acts on $2 \times 2$ spin space
and $P$ represents the space inversion operator.
If the system is square shaped with $N \times N$ sites
(i.e., $1 \le m, n \le N$), then $P$ is expressed as
\begin{align}
  P = \sum_{m,n}|N+1-m,N+1-n \rangle \langle m,n| .
\end{align}
From this symmetry, we can show that, if $|\Psi^{R}\rangle$ satisfies
$H|\Psi^{R}\rangle = E |\Psi^{R}\rangle$,
then ${}^{t}(UP|\Psi^{R}\rangle)$ satisfies
${}^{t}(UP|\Psi^{R}\rangle)H = {}^{t}(UP|\Psi^{R}\rangle)(-E)$.
That is, an eigenstate of $H$ with an eigenvalue of $E$ is paired with
its counterpart with the eigenvalue of $-E$ if $|E| \neq 0$.

These results tell us that if an eigenstate with an eigenvalue
of $E$ with ${\rm Re}\{E\} \neq 0$ and ${\rm Im}\{E\} \neq 0$ is present,
this eigenstate and its three counterparts form a quartet:
four eigenstates with $E$, $E^{*}$, $-E^{*}$, and $-E$ are related
by the two symmetries.
An important exception is the case of ${\rm Re}\{E\} \neq 0$ and
${\rm Im}\{E\} = 0$, where an eigenstate and its counterpart
form a pair; eigenstates with $E$ and $-E$ are related by the symmetries.

The above statement is useful in considering a complex spectrum of our system.
Let us denote a complex eigenvalue of energy $E = E_{\rm R}+iE_{\rm I}$
as $E = (E_{\rm R}, E_{\rm I})$.
All eigenvalues are real as $E = (\varepsilon, 0)$
in the Hermitian limit of $\gamma = 0$.
With increasing $\gamma$, eigenvalues gradually become complex
as $E = (\varepsilon, \delta$).
We focus on a pair of eigenstates with eigenvalues of
$(\varepsilon_{1}, 0)$ and $(-\varepsilon_{1}, 0)$ at $\gamma = 0$.
Let us suppose that $(\varepsilon_{1}, 0)$ deviates from the real axis
as $(\varepsilon_{1}, \delta)$ with increasing $\gamma$.
This statement requires that this eigenstate forms a quartet
together with three other eigenstates with eigenvalues of
$(\varepsilon_{1}, -\delta)$, $(-\varepsilon_{1}, \delta)$,
and $(-\varepsilon_{1}, -\delta)$.
Such a quartet should be formed by
a combination of two pairs of eigenstates.
That is, an eigenvalue can become complex only when a pair of eigenstates
with $(\varepsilon_{1}, 0)$ and $(-\varepsilon_{1}, 0)$
and another pair of eigenstates with $(\varepsilon_{2}, 0)$ and
$(-\varepsilon_{2}, 0)$ are degenerate in energy
under the condition of $\varepsilon_{1} = \varepsilon_{2}$.

\section{Right and Left Eigenstates}

We consider the system of $N \times N$ sites
in the boundary and bulk geometries.
Right and left eigenstates of $H$ are expressed as
\begin{align}
 &  |\Psi^{R}\rangle
    = \sum_{m,n} |m,n \rangle \cdot |\psi^{R}(m,n)\rangle ,
     \\
 &  \langle\Psi^{L}|
    = \sum_{m,n} \langle\psi^{L}(m,n)| \cdot \langle m,n| ,
\end{align}
where $|\psi^{R}(m,n)\rangle$ and $\langle\psi^{L}(m,n)|$ are, respectively,
two component column and row vectors:
\begin{align}
  |\psi^{R}(m,n)\rangle
  & = \left[ \begin{array}{c}
               \psi_{\uparrow}^{R}(m,n) \\
               \psi_{\downarrow}^{R}(m,n)
             \end{array}
      \right] ,
     \\
  \langle\psi^{L}(m,n)|
  & = \left[ \begin{array}{cc}
               \psi_{\uparrow}^{L}(m,n) & \psi_{\downarrow}^{L}(m,n)
             \end{array}
      \right] .
\end{align}
In the boundary geometry, the following obc is imposed on
$|\psi^{R}(m,n)\rangle$ and $\langle\psi^{L}(m,n)|$:
\begin{align}
    \label{eq:obc^R}
 & |\psi^{R}(0,n)\rangle = |\psi^{R}(N+1,n)\rangle
       \nonumber \\
 & = |\psi^{R}(m,0)\rangle = |\psi^{R}(m,N+1)\rangle
   = \left[ \begin{array}{c}
               0 \\
               0
            \end{array}
     \right] ,
        \\
    \label{eq:obc^L}
 & \langle\psi^{L}(0,n)| = \langle\psi^{L}(N+1,n)|
       \nonumber \\
 & = \langle\psi^{L}(m,0)| = \langle\psi^{L}(m,N+1)|
   = \left[ \begin{array}{cc}
               0 & 0
            \end{array}
     \right] .
\end{align}

In the bulk geometry, we introduce plane-wave-like right and left eigenstates
to define a topological number.
To do this, we set
\begin{align}
    \label{eq:psi^R}
  |\psi^{R}(m,n)\rangle
  & = \varphi^{R}(m,n) |\psi^{R}\rangle ,
      \\
    \label{eq:psi^L}
  \langle\psi^{L}(m,n)|
  & = \langle \psi^{L}| \varphi^{L}(m,n) ,
\end{align}
where $|\psi\rangle^{R}$ and $\langle\psi^{L}|$ are, respectively,
two component column and row vectors independent of $m$ and $n$, and
\begin{align}
  \varphi^{R}(m,n) & = \beta_{x}^{m}\beta_{y}^{n} ,
    \\
  \varphi^{L}(m,n) & = {\beta'}_{\!\!x}^{m}{\beta'}_{\!\!y}^{n} .
\end{align}
The eigenvalue equation $H|\Psi^{R}\rangle = E|\Psi^{R}\rangle$
is reduced to
\begin{align}
  H_{\rm rd}^{R}|\psi^{R}\rangle = E|\psi^{R}\rangle ,
\end{align}
where $H_{\rm rd}^{R} = \sigma_{x}\eta_{x} + \sigma_{y}\eta_{y}
+ \sigma_{z}\eta_{z}$ with
\begin{align}
     \label{eq:def-eta_x}
 & \eta_{x}
   = \frac{1}{2i}\left(\beta_{x}-\beta_{x}^{-1}\right)+i\gamma ,
      \\
     \label{eq:def-eta_y}
 & \eta_{y}
   = \frac{1}{2i}\left(\beta_{y}-\beta_{y}^{-1}\right)+i\gamma ,
      \\
     \label{eq:def-eta_z}
 & \eta_{z}
   = M - \frac{1}{2}\left(\beta_{x}+\beta_{x}^{-1}\right)
       - \frac{1}{2}\left(\beta_{y}+\beta_{y}^{-1}\right) .
\end{align}
The eigenvalue equation $\langle \Psi^{L}|H = \langle \Psi^{L}|E$
is reduced to
\begin{align}
  \langle \psi^{L}| H_{\rm rd}^{L} = \langle \psi^{L}|E ,
\end{align}
where $H_{\rm rd}^{L} = \sigma_{x}{\eta'}_{\!\!x}+\sigma_{y}{\eta'}_{\!\!y}
+ \sigma_{z}{\eta'}_{\!\!z}$ with
\begin{align}
 & {\eta'}_{\!\!x}
   = -\frac{1}{2i}\left({\beta'}_{\!\!x}-{\beta'}_{\!\!x}^{-1}\right)
     +i\gamma ,
      \\
 & {\eta'}_{\!\!y}
   = -\frac{1}{2i}\left({\beta'}_{\!\!y}-{\beta'}_{\!\!y}^{-1}\right)
     +i\gamma ,
      \\
 & {\eta'}_{\!\!z}
   = M - \frac{1}{2}\left({\beta'}_{\!\!x}+{\beta'}_{\!\!x}^{-1}\right)
       - \frac{1}{2}\left({\beta'}_{\!\!y}+{\beta'}_{\!\!y}^{-1}\right) .
\end{align}
We impose the following mpbc on $\varphi^{R}(m,n)$:
\begin{align}
      \label{eq:BC^R_mpbc}
 \varphi^{R}(m+N,n) = \varphi^{R}(m,n+N) = b^{N}\varphi^{R}(m,n) ,
\end{align}
which results in
\begin{align}
 & \beta_{x} = be^{ik_{x}a},
   \hspace{3mm} \beta_{y} = be^{ik_{y}a} ,
\end{align}
where $k_{i} = 2n_{i}\pi/N$ and $n_{i} = 0,1,2,\dots, N-1$ ($i = x, y$).
To present a biorthogonal set of right and left basis functions,~\cite{brody}
we determine ${\beta'}_{\!\!x}$ and ${\beta'}_{\!\!y}$ as
\begin{align}
     \label{eq:beta'}
   {\beta'}_{\!\!x} = \beta_{x}^{-1}, \hspace{10mm}
   {\beta'}_{\!\!y} = \beta_{y}^{-1} ,
\end{align}
which results in
\begin{align}
 & {\beta'}_{\!\!x} = (be^{ik_{x}a})^{-1} ,
   \hspace{3mm} {\beta'}_{\!\!y} = (be^{ik_{y}a})^{-1} .
\end{align}
This indicates that $\varphi^{L}(m,n)$ obeys the following mpbc:
\begin{align}
      \label{eq:BC^L_mpbc}
 \varphi^{L}(m+N,n) = \varphi^{L}(m,n+N) = b^{-N}\varphi^{L}(m,n) ,
\end{align}
which is different from Eq.~(\ref{eq:BC^R_mpbc}).

We find $H_{\rm rd}^{L} = H_{\rm rd}^{R}$ from Eq.~(\ref{eq:beta'})
and thus denote them as $H_{\rm rd}$ hereafter.
The reduced Hamiltonian $H_{\rm rd}$ is expressed as
\begin{align}
   H_{\rm rd}
   = \sigma_{x}\eta_{x} + \sigma_{y}\eta_{y} + \sigma_{z}\eta_{z} ,
\end{align}
with
\begin{align}
     \label{eq:def-eta_x-mod}
 & \eta_{x}
   = b_{+}\sin(k_{x}a) +i\left[ \gamma - b_{-}\cos(k_{x}a) \right] ,
      \\
 & \eta_{y}
  = b_{+}\sin(k_{y}a) +i\left[ \gamma - b_{-}\cos(k_{y}a) \right] ,
      \\
 & \eta_{z}
   = M - b_{+}\left[\cos(k_{x}a)+\cos(k_{y}a)\right]
      \nonumber \\
 & \hspace{20mm}
       -ib_{-}\left[ \sin(k_{x}a) + \sin(k_{x}a) \right] .
\end{align}
where
\begin{align}
    b_{\pm} = \frac{1}{2}\left(b \pm b^{-1}\right) .
\end{align}

By solving the reduced eigenvalue equation, we find that the energy of
an eigenstate characterized by $\mib{k}=(k_{x},k_{y})$ and $b$
is given by $E = \pm \epsilon$ with
\begin{align}
    \label{eq:def-epsilon}
 \epsilon = \sqrt{\eta_{x}^{2}+\eta_{y}^{2}+\eta_{z}^{2}} ,
\end{align}
where ${\rm Re}\{\epsilon\} \ge 0$.
The right and left eigenstates of $H$ with the eigenvalue of $\pm\epsilon$
are expressed as
\begin{align}
    |\Psi_{\mib{k}}^{R\pm}\rangle
  & = \frac{1}{N} \sum_{m,n}
      (be^{ik_{x}a})^{m}(be^{ik_{y}a})^{n} |m,n \rangle
      \cdot |\psi_{\mib{k}}^{R\pm}\rangle ,
     \\
    \langle \Psi_{\mib{k}}^{L\pm}|
  & = \frac{1}{N} \sum_{m,n}
      \langle\psi_{\mib{k}}^{L\pm}| \cdot \langle m,n|
      (be^{ik_{x}a})^{-m}(be^{ik_{y}a})^{-n} ,
\end{align}
where
\begin{align}
       \label{eq:psi_k_R}
  |\psi_{\mib{k}}^{R\pm}\rangle
  & = \frac{1}{\sqrt{A_{\pm}}}
                    \left[ \begin{array}{c}
                              \eta_{z}\pm\epsilon \\
                              \eta_{x}+i\eta_{y}
                     \end{array}
                    \right] ,
      \\
       \label{eq:psi_k_L}
  \langle\psi_{\mib{k}}^{L\pm}|
  & = \frac{1}{\sqrt{A_{\pm}}}
             ^{t}\! \left[ \begin{array}{c}
                              \eta_{z}\pm\epsilon \\
                              \eta_{x}-i\eta_{y}
                     \end{array}
                    \right] ,
\end{align}
with $A_{\pm} = \eta_{x}^{2}+\eta_{y}^{2}+(\eta_{z}\pm\epsilon)^{2}$.
We easily find
\begin{align}
  \langle \Psi_{\mib{k}}^{L\pm}|\Psi_{\mib{k'}}^{R\pm}\rangle
  = \delta_{\mib{k},\mib{k'}} ,
  \hspace{6mm}
  \langle \Psi_{\mib{k}}^{L\pm}|\Psi_{\mib{k'}}^{R\mp}\rangle
  = 0 .
\end{align}
That is, $\{|\Psi_{\mib{k}}^{R\pm}\rangle\}$
and $\{\langle \Psi_{\mib{k}}^{L\pm}|\}$
constitute a biorthogonal set of basis functions.~\cite{brody}
The eigenvectors satisfy
\begin{align}
 \langle\psi_{\mib{k}}^{L\pm}|\psi_{\mib{k}}^{R\pm}\rangle = 1 ,
 \hspace{6mm}
 \langle\psi_{\mib{k}}^{L\pm}|\psi_{\mib{k}}^{R\mp}\rangle = 0 .
\end{align}

\section{Chiral Edge States}

We determine the spectrum of chiral edge states in the $N \times N$ system
when an obc is imposed in one direction
and a mpbc is imposed in the other direction.
Considering the resulting spectrum, we find important characteristics of
chiral edge states in the boundary geometry.

First, we consider the system [see Fig.~1(a)]
under the obc in the $y$-direction,
\begin{align}
  \varphi^{R}(m,0) & = \varphi^{R}(m,N+1) = 0 ,
\end{align}
and the mpbc in the $x$-direction,
\begin{align}
  \varphi^{R}(m,n) & = b^{N} \varphi^{R}(m+N,n) .
\end{align}
To obtain fundamental solutions, we set
\begin{align}
  \varphi^{R}(m,n) & = \beta_{x}^{m}\rho^{n} ,
\end{align}
where $\rho$ characterizes the penetration of chiral edge states
in the $y$-direction and $\beta_{x} = be^{ik_{x}a}$.
We decompose the reduced Hamiltonian as
$H_{\rm rd} = H_{\rm rd}^{(0)} + H_{\rm rd}^{(1)}$ with
\begin{align}
   H_{\rm rd}^{(0)}
 & = i\left[ \gamma-\frac{\rho-\rho^{-1}}{2} \right]
       \sigma_{y}
     + \left[ \tilde{M}-\frac{\rho+\rho^{-1}}{2} \right]
       \sigma_{z} ,
           \\
   H_{\rm rd}^{(1)}
 & = \eta_{x}\sigma_{x} ,
\end{align}
where $\eta_{x}$ is given in Eq.~(\ref{eq:def-eta_x-mod}) and
\begin{align}
   \tilde{M} = M - \frac{1}{2}\left(\beta_{x}+\beta_{x}^{-1}\right) .
\end{align}

A simple method for obtaining edge localized solutions is
described in Appendix of Ref.~\citen{takane2}.
By adapting the method to
$H_{\rm rd}^{(0)}|\psi^{R}\rangle = E_{\bot}|\psi^{R}\rangle$,
we find two eigenstates at $E_{\bot} = 0$:
\begin{align}
      \label{eq:Psi_1-bottom}
    |\Psi_{1}^{R}\rangle
  & = c \sum_{m}(be^{ik_{x}a})^{m}
      \sum_{n=0}^{N+1}
      \left( \rho_{1}^{n} - \tilde{\rho}_{1}^{n} \right)
        \nonumber \\
  & \hspace{2mm} \times 
      |m,n \rangle \cdot |\psi_{1}^{R}\rangle ,
     \\
      \label{eq:Psi_2-upper}
    |\Psi_{2}^{R}\rangle
  & = c \sum_{m}(be^{ik_{x}a})^{m}
      \sum_{n=0}^{N+1}
      \left( \frac{\rho_{2}^{n}}{\rho_{2}^{N+1}}
             - \frac{\tilde{\rho}_{2}^{n}}{\tilde{\rho}_{2}^{N+1}} \right)
        \nonumber \\
  & \hspace{2mm} \times
      |m,n \rangle \cdot |\psi_{2}^{R}\rangle ,
\end{align}
where $c$ is a normalization constant,
\begin{align}
  |\psi_{1}^{R}\rangle
  = \frac{1}{\sqrt{2}}
                    \left[ \begin{array}{c}
                              1 \\
                              1
                     \end{array}
                    \right] ,
  \hspace{5mm}
  |\psi_{2}^{R}\rangle
  = \frac{1}{\sqrt{2}}
                    \left[ \begin{array}{c}
                              1 \\
                              -1
                     \end{array}
                    \right] ,
\end{align}
and
\begin{align}
     \label{eq:rho-1,2}
  \rho_{1} = \tilde{M}+\gamma, \hspace{5mm}
  \rho_{2} = \frac{1}{\tilde{M}-\gamma} .
\end{align}
The limit of $\tilde{\rho}_{1} \to 0$ and $\tilde{\rho}_{2} \to \infty$
is assumed in the above equations, so that
\begin{align}
  \sum_{n=0}^{N+1}\tilde{\rho}_{1}^{n}|m,n \rangle
  = |m,0 \rangle ,
  \hspace{2mm}
  \sum_{n=0}^{N+1}\frac{\tilde{\rho}_{2}^{n}}{\tilde{\rho}_{2}^{N+1}}
  |m,n \rangle
  = |m,N+1 \rangle .
\end{align}
We observe that $|\Psi_{1}^{R}\rangle$ satisfies the obc at $n = 0$.
When $|\rho_{1}| < 1$, it exponentially decreases with increasing $n$
and satisfies the obc at $n = N+1$ if $N$ is sufficiently large.
That is, $|\Psi_{1}^{R}\rangle$ is localized near the bottom edge at $n = 1$
when $|\rho_{1}| < 1$.
In a manner similar to this, we can show that $|\Psi_{2}^{R}\rangle$
is localized near the top edge at $n = N$ when $1< |\rho_{2}|$.
Since $H_{\rm rd}^{(1)}|\psi_{1}^{R}\rangle = \eta_{x}|\psi_{1}^{R}\rangle$
and $H_{\rm rd}^{(1)}|\psi_{2}^{R}\rangle = -\eta_{x}|\psi_{2}^{R}\rangle$,
we can conclude that the energy dispersion relations of the edge states
localized near the bottom edge and top edge respectively are
\begin{align}
      \label{eq:disp-b}
 & E_{\rm bottom}(k_{x})
    = b_{+}\sin(k_{x}a) +i\left[ \gamma -b_{-}\cos(k_{x}a) \right] ,
        \\
      \label{eq:disp-u}
 & E_{\rm top}({k^{\prime}}_{\!\!x})
    = -b_{+}\sin({k^{\prime}}_{\!\!x}a)
      -i\left[ \gamma -b_{-}\cos({k^{\prime}}_{\!\!x}a) \right] ,
\end{align}
where $k_{x}$ in $E_{\rm top}$ is replaced with ${k^{\prime}}_{\!\!x}$
for later convenience.
%%%%%%%%%%%%%%%%%%
\begin{figure}[btp]
\begin{center}
\includegraphics[height=2.8cm]{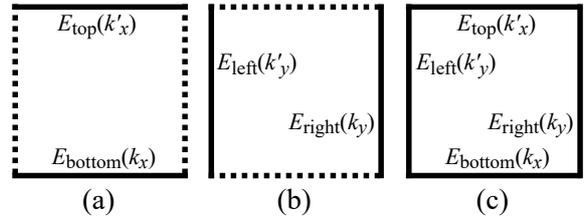}
\end{center}
\caption{
Three square geometries used to consider chiral edge states.
The obc is imposed on edges denoted by solid lines
and the mpbc is imposed on a pair of edges denoted by dotted lines.
}
\end{figure}
%%%%%%%%%%%%%%%%%%

We next consider the system  [see Fig.~1(b)]
under the obc in the $x$-direction,
\begin{align}
  \varphi^{R}(0,n) & = \varphi^{R}(N+1,n) = 0 ,
\end{align}
and the mpbc in the $y$-direction,
\begin{align}
  \varphi^{R}(m,n) & = b^{N} \varphi^{R}(m,n+N) ,
\end{align}
and find that the energy dispersion relations of the edge states
localized near the left edge and right edge respectively are
\begin{align}
      \label{eq:disp-l}
 & E_{\rm left}({k^{\prime}}_{\!\!y})
    = -b_{+}\sin({k^{\prime}}_{\!\!y}a)
      -i\left[ \gamma -b_{-}\cos({k^{\prime}}_{\!\!y}a) \right] ,
      \\
      \label{eq:disp-r}
 & E_{\rm right}(k_{y})
    = b_{+}\sin(k_{y}a) +i\left[ \gamma -b_{-}\cos(k_{y}a) \right] ,
\end{align}
where $k_{y}$ in $E_{\rm left}$ is replaced with ${k^{\prime}}_{\!\!y}$
for later convenience.

In the above argument, we assume $|\rho_{1}| < 1 < |\rho_{2}|$
such that two solutions are spatially separated:
one is localized near the bottom (right) edge
whereas the other is localized near the top (left) edge.
Although this assumption is satisfied when $\gamma$ is sufficiently small,
it eventually breaks down with increasing $\gamma$.
However, as briefly described in Appendix, even though the assumption
breaks down, the dispersion relations in Eqs.~(\ref{eq:disp-b}),
(\ref{eq:disp-u}), (\ref{eq:disp-l}), and (\ref{eq:disp-r}) are valid
as long as $N$ is sufficiently large and $|\rho_{1}| < |\rho_{2}|$.
Here, $|\rho_{1}| < |\rho_{2}|$ is satisfied
under the condition of $M < 2\sqrt{1+\gamma^{2}}$ if $\gamma < 1$.
In Sect.~5, we show that this condition describes
one part of phase boundaries.

Let us consider chiral edge states
in the boundary geometry shown in Fig.~1(c).
A chiral edge state circulates along the square loop
consisting of the bottom, top, left, and right edges.
Thus, we expect that such a chiral edge state can be allowed only when
the four dispersion relations in Eqs.~(\ref{eq:disp-b}), (\ref{eq:disp-u}),
(\ref{eq:disp-l}), and (\ref{eq:disp-r}) become identical for appropriately
chosen $k_{x}$, ${k^{\prime}}_{\!\!x}$, ${k^{\prime}}_{\!\!y}$, and $k_{y}$:
\begin{align}
  E_{\rm bottom}(k_{x}) = E_{\rm top}({k^{\prime}}_{\!\!x})
  = E_{\rm left}({k^{\prime}}_{\!\!y})=E_{\rm right}(k_{y}) .
\end{align}
This is satisfied when
$k_{x} = -{k^{\prime}}_{\!\!x} = -{k^{\prime}}_{\!\!y} = k_{y} \equiv k$
with
\begin{align}
    \label{eq:condition-real}
  b_{-}\cos(ka) = \gamma .
\end{align}
Note that the mpbc is essential in obtaining this result.

Equation~(\ref{eq:condition-real}) requires that the energy $E$ of
a chiral edge state must be real.
This result is consistent with the argument based on the symmetries of $H$.
In Sect.~2, we show that an eigenvalue of energy can become complex only when
a pair of eigenstates with $(\varepsilon_{1}, 0)$ and $(-\varepsilon_{1}, 0)$
and another pair of eigenstates with $(\varepsilon_{2}, 0)$ and
$(-\varepsilon_{2}, 0)$ are degenerate in energy
under the condition of $\varepsilon_{1} = \varepsilon_{2}$.
Let us apply this to the chiral edge states
in the Hermitian limit of $\gamma = 0$, where its spectrum is
almost equally distributed on the real axis of $E$ in a pairwise manner.
Even when $\gamma$ is increased from $0$, two pairs of chiral edge states
cannot be degenerate when the chiral edge states are topologically
protected and the spectrum is almost equally distributed on the real axis.
Therefore, the spectrum of the chiral edge states cannot deviate
from the real axis as long as the states are stable.
Conversely, we can say that a chiral edge state is destabilized	
if its energy becomes complex as $E = (\varepsilon, \delta)$.

In the boundary geometry, Eq.~(\ref{eq:condition-real}) also requires
that the wavefunction amplitude of the chiral edge states varies
in the $x$- and $y$-directions with the same rate of increase:
\begin{align}
     \label{eq:b-CEs_bg}
  b = \sqrt{\left(\frac{\gamma}{\cos (ka)}\right)^{2}+1}
      + \frac{\gamma}{\cos (ka)} .
\end{align}
This result is used in the argument in Sect.~5.

\section{Bulk--Boundary Correspondence}

In Sect.~3, it is shown that the plane-wave-like states under the mpbc
of Eq.~(\ref{eq:BC^R_mpbc}) consist of
a conduction band of $E = \epsilon$ and a valence band of $E = -\epsilon$,
where $\epsilon$ is given in Eq.~(\ref{eq:def-epsilon}).
When the two bands are separated by a gap,
that is, ${\rm Re}\{\epsilon\} \neq 0$ for an arbitrary $\mib{k}$,
we can define the Chern number $\nu$
by using $|\psi_{\mib{k}}^{R-}\rangle$ and $\langle\psi_{\mib{k}}^{L-}|$
given in Eqs.~(\ref{eq:psi_k_R}) and (\ref{eq:psi_k_L}).
The result is~\cite{yao2}
\begin{align}
          \label{eq:def-N}
  \nu = \frac{1}{2\pi i}\int d^{2}k
        \left( \frac{\partial \langle\psi_{\mib{k}}^{L-}|}{\partial k_{x}}
               \frac{\partial |\psi_{\mib{k}}^{R-}\rangle}{\partial k_{y}}
             - \frac{\partial \langle\psi_{\mib{k}}^{L-}|}{\partial k_{y}}
               \frac{\partial |\psi_{\mib{k}}^{R-}\rangle}{\partial k_{x}}
        \right) ,
\end{align}
which is rewritten as~\cite{takane1}
\begin{align}
  \nu = \int \frac{d^{2}k}{4 \pi} \frac{1}{\epsilon^{3}}
        \left(  \eta_{x}\frac{\partial \eta_{y}}{\partial k_{y}}
                \frac{\partial \eta_{z}}{\partial k_{x}}
              + \frac{\partial \eta_{x}}{\partial k_{x}}
                \eta_{y}\frac{\partial \eta_{z}}{\partial k_{y}}
              - \frac{\partial \eta_{x}}{\partial k_{x}}
                \frac{\partial \eta_{y}}{\partial k_{y}}\eta_{z}
        \right) .
\end{align}
Here, the limit of $N \to \infty$ is implicitly assumed.
The Chern number $\nu$ takes an integer that depends on
not only $M$ and $\gamma$ but also $b$.
In the Hermitian limit of $\gamma = 0$,
Eq.~(\ref{eq:def-N}) at $b = 1$ becomes equivalent to
an ordinary expression of the Chern number.~\cite{thouless,kohmoto}

Our system takes three phases: a topologically trivial phase of $\nu = 0$,
a nontrivial phase of $\nu = 1$,
and a gapless phase, in which $\nu$ cannot be defined.
We consider $\nu$ for a given $M$ in a parameter space of $\gamma$ and $b$,
where these phases are separated by lines on which the gap closes.
Let us find such gap closing lines.~\cite{takane1}
A gap closing takes place in our system when
${\rm Re}\{\epsilon\} = {\rm Im}\{\epsilon\} = 0$.
Equation~(\ref{eq:def-epsilon}) indicates that
${\rm Im}\{\epsilon^{2}\} = 0$ for
$\mib{k} =(0,0)$, $(0,\frac{\pi}{a})$, $(\frac{\pi}{a},0)$,
and $(\frac{\pi}{a},\frac{\pi}{a})$,
where $\mib{k} = (\frac{\pi}{a},\frac{\pi}{a})$ can be ignored
because this point has a gap in relevant situations.
For $\mib{k}=(0,0)$, we find that ${\rm Re}\{\epsilon^{2}\} = 0$ when
$(M-2b_{+})^{2} - 2(-b_{-}+\gamma)^{2} = 0$,
which yields four solutions of $b$:
\begin{align}
      \label{eq:b_1}
  b_{\rm 1} = \frac{M-\sqrt{2}\gamma
                    + \sqrt{(M-\sqrt{2}\gamma)^{2}-2}}
                   {2-\sqrt{2}} ,
         \\
      \label{eq:b_2}
  b_{\rm 2} = \frac{M+\sqrt{2}\gamma
                    + \sqrt{(M+\sqrt{2}\gamma)^{2}-2}}
                   {2+\sqrt{2}} ,
         \\
      \label{eq:b_3}
  b_{\rm 3} = \frac{M-\sqrt{2}\gamma
                    - \sqrt{(M-\sqrt{2}\gamma)^{2}-2}}
                   {2-\sqrt{2}} ,
         \\
      \label{eq:b_4}
  b_{\rm 4} = \frac{M+\sqrt{2}\gamma
                    - \sqrt{(M+\sqrt{2}\gamma)^{2}-2}}
                   {2+\sqrt{2}} .
\end{align}
For $\mib{k} =(0,\frac{\pi}{a})$ and $(\frac{\pi}{a},0)$,
we find that ${\rm Re}\{\epsilon^{2}\} = 0$ when
$2b_{-}^{2} -M^{2} + 2\gamma^{2} = 0$,
which yields two solutions of $b$:
\begin{align}
      \label{eq:b_5}
  b_{5} & =  \sqrt{\frac{M^{2}}{2}-\gamma^{2}+1}
             + \sqrt{\frac{M^{2}}{2}-\gamma^{2}} ,
         \\
      \label{eq:b_6}
  b_{6} & = \sqrt{\frac{M^{2}}{2}-\gamma^{2}+1}
            - \sqrt{\frac{M^{2}}{2}-\gamma^{2}} .
\end{align}
Each $b_{i}$ ($i = 1, \dots, 6$) is irrelevant if it becomes a complex number.
In addition to $b_{i}$ ($i = 1, \dots, 6$),
we need to consider $b_{0}$ given below.
When $(M-\sqrt{2}\gamma)^{2} < 2$, $b_{1}$ and $b_{3}$
become the complex numbers $b_{0}e^{\pm ik_{0}a}$, where
\begin{align}
    \label{eq:b_0}
  b_{0} = 1+\sqrt{2}
\end{align}
and $k_{0}$ is defined by
\begin{align}
  \tan\left(k_{0}a\right)
   = \frac{\sqrt{2-(M-\sqrt{2}\gamma)^{2}}}
                    {M-\sqrt{2}\gamma} .
\end{align}
We can show that ${\rm Re}\{\epsilon^{2}\} = {\rm Im}\{\epsilon^{2}\} = 0$
at $b = b_{0}$ if $\mib{k}=\pm (k_{0},k_{0})$.
Therefore, $b_{0}$ is a gap closing line.

The phase diagrams in the bulk geometry for $M = 1.0$, $2.0$,
$2.5$, and $3.0$ are shown in Figs.~2(a)--2(d) in the $\gamma b$-plane,
where the gap closing lines separate
the topologically trivial region ($\nu = 0$),
nontrivial region ($\nu = 1$),
and gapless region in which $\nu$ cannot be defined.
For example, the nontrivial region ($\nu = 1$) in panel~(a) is
bounded by $b_{2}$, $b_{5}$, and $b_{6}$,
and its outside is the gapless region.
%%%%%%%%%%%%%%%%%%
\begin{figure}[btp]

\begin{tabular}{cc}
\begin{minipage}{0.5\hsize}
\begin{center}
\hspace{-10mm}
\includegraphics[height=4.2cm]{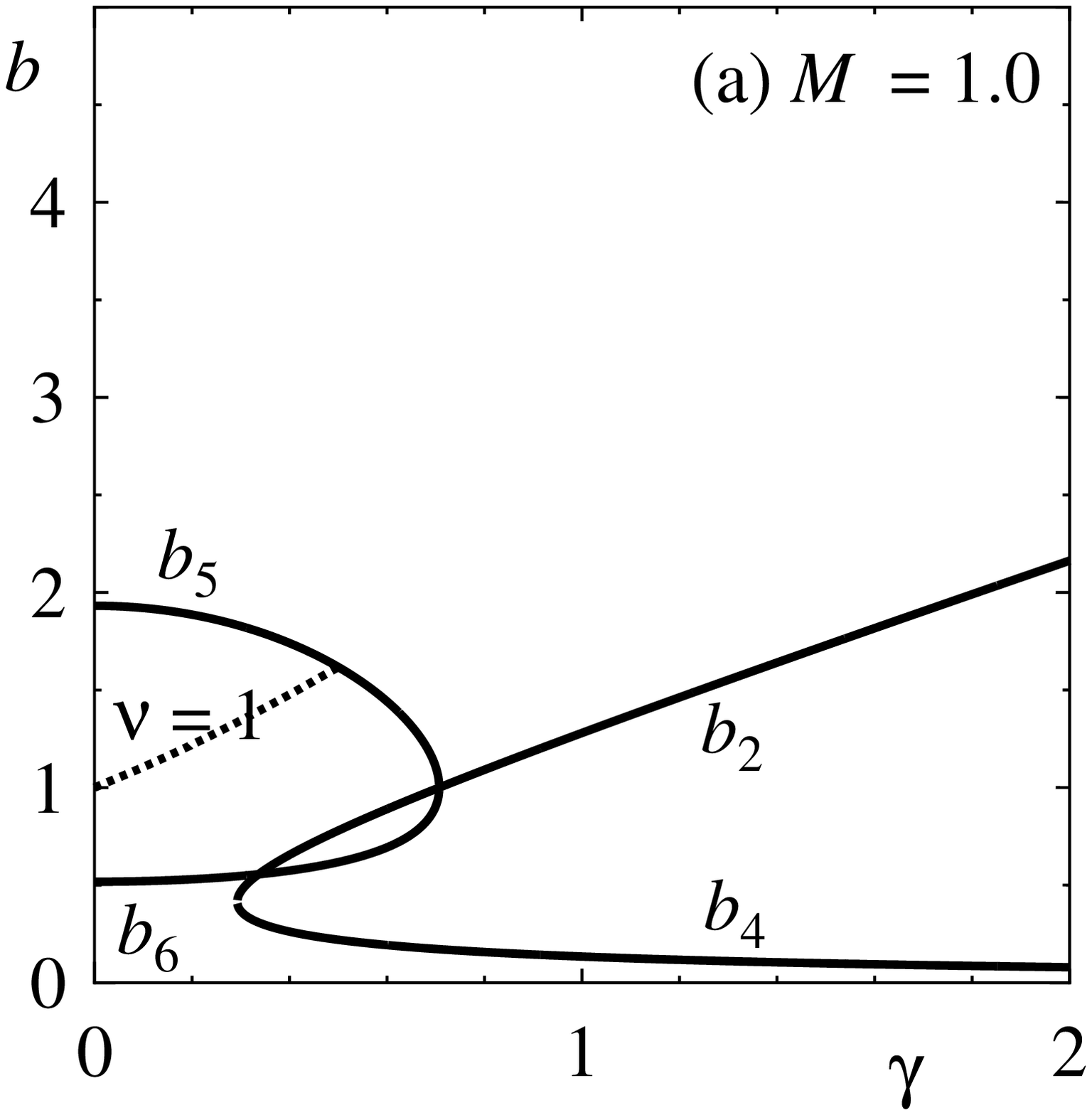}
\end{center}
\end{minipage}
\begin{minipage}{0.5\hsize}
\begin{center}
\hspace{-10mm}
\includegraphics[height=4.2cm]{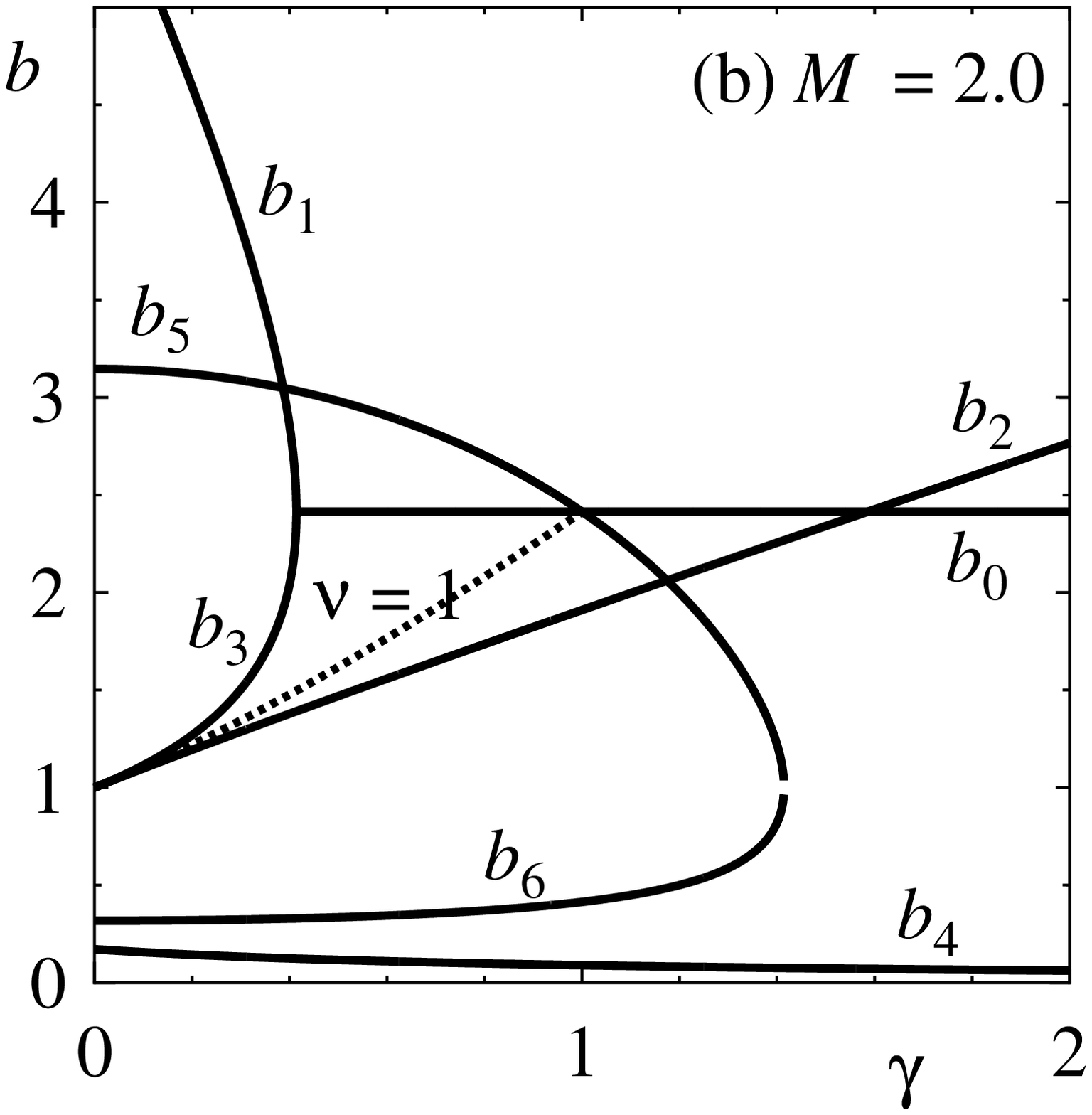}
\end{center}
\end{minipage}
\end{tabular}

\begin{tabular}{cc}
\begin{minipage}{0.5\hsize}
\begin{center}
\hspace{-10mm}
\includegraphics[height=4.2cm]{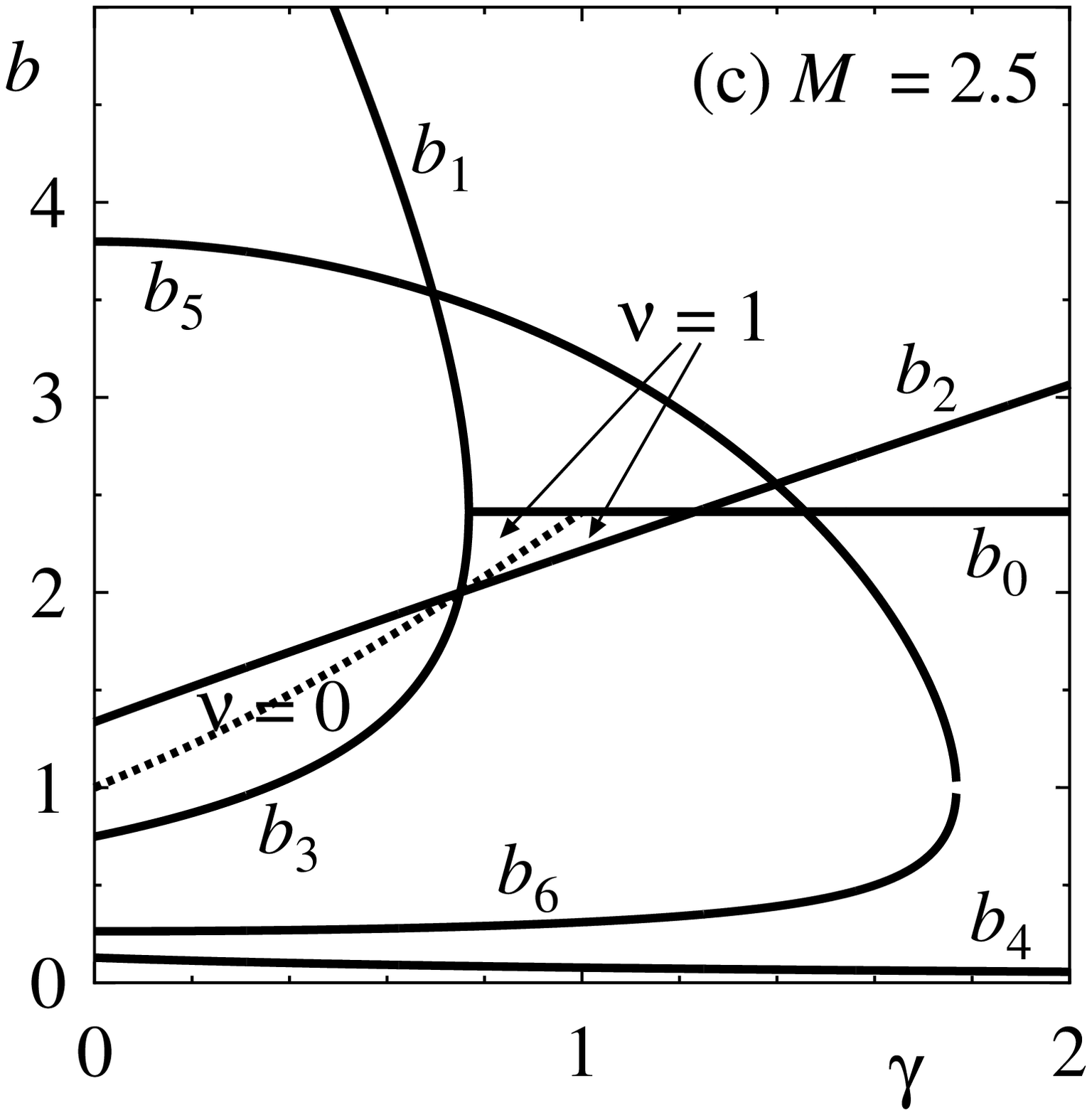}
\end{center}
\end{minipage}
\begin{minipage}{0.5\hsize}
\begin{center}
\hspace{-10mm}
\includegraphics[height=4.2cm]{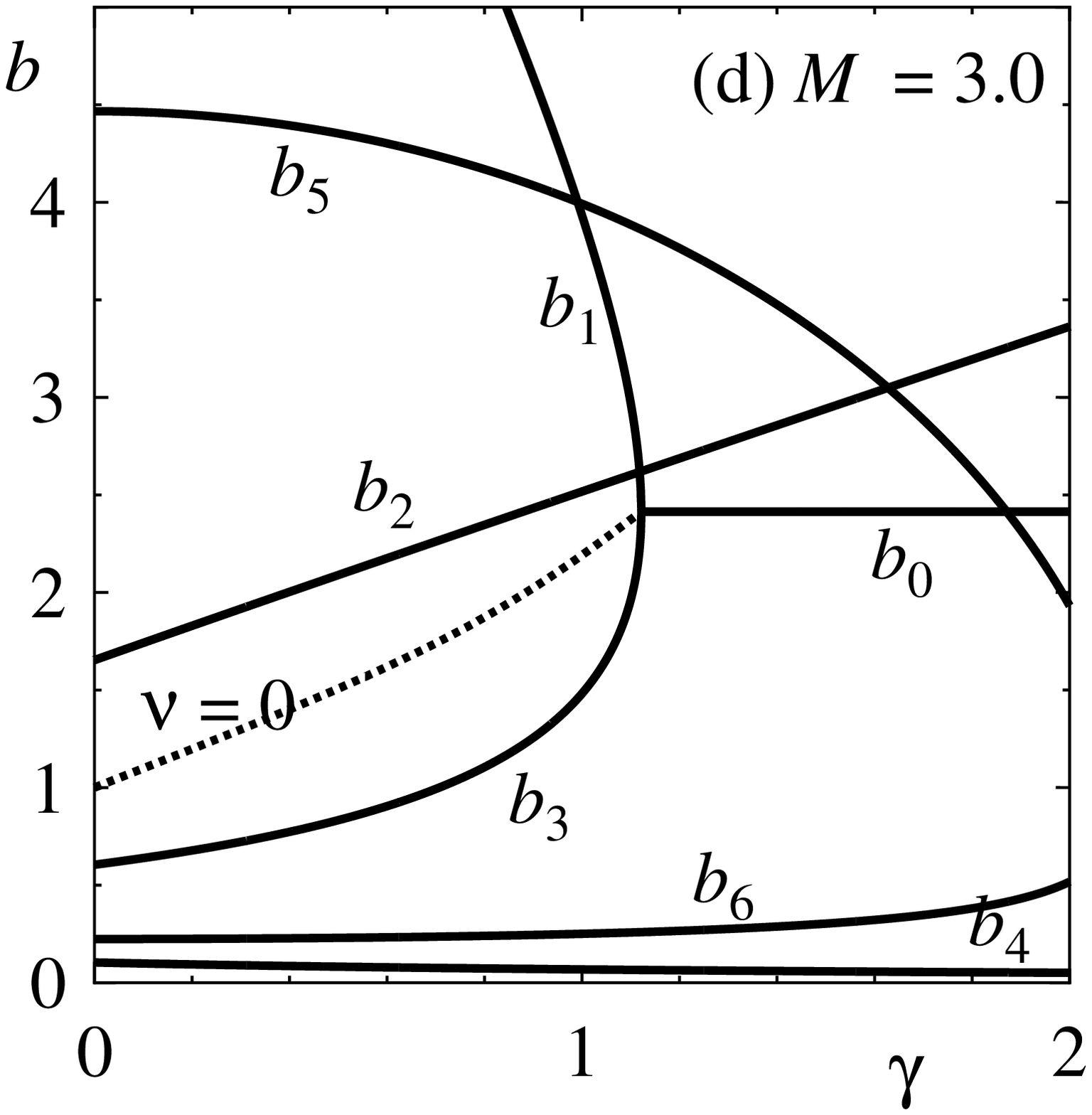}
\end{center}
\end{minipage}
\end{tabular}
\caption{
Phase diagrams in the bulk geometry in the $\gamma b$-plane for
$M =$ (a) $1.0$, (b) $2.0$, (c) $2.5$, and (d) $3.0$.
Topologically trivial and nontrivial regions are respectively
designated as $\nu =0$ and $\nu =1$, and the outside of these regions is
a gapless region, in which $\nu$ cannot be defined.
In each panel, the dotted line represents the trajectory of
$b(\gamma)$.
}
\end{figure}
%%%%%%%%%%%%%%%%%%

Using the phase diagrams shown in Fig.~2,
let us consider which of the three phases appears in the boundary geometry.
In the Hermitian limit of $\gamma = 0$, a phase realized
in the boundary geometry is governed by $\nu$
at $b = 1$, which corresponds to the ordinary pbc.
If $\nu = 1$ ($\nu = 0$) at $b = 1$, the nontrivial (trivial) phase
is realized in the boundary geometry.
To extend this bulk--boundary correspondence to the non-Hermitian regime
of $\gamma > 0$, we need to determine $b$ as a function of
$\gamma$ such that $\nu(\gamma,b)$ is
in one-to-one correspondence with a phase realized in the boundary geometry.

A recipe for determining $b(\gamma)$ is given in Ref.~\citen{takane1}.
We here give a revised version of it:
\begin{enumerate}
\item $b(0) = 1$ at the Hermitian limit of $\gamma = 0$.
\item With the exception described in (3), $b(\gamma)$ is allowed to cross
gap closing lines only at a crossing point between the two.
\item If a peculiar gap closing is caused by the destabilization of
topological boundary states, $b(\gamma)$ in the nontrivial region
is determined in accordance with it.
\end{enumerate}
The third requirement is added in this revised version.
Below, we explain the second and third requirements as well as
the meaning of a peculiar gap closing.

We briefly describe the reasoning
on which the second requirement is based.~\cite{takane1}
If $b(\gamma)$ crosses a gap closing line, a zero-energy solution appears
at the crossing point, giving rise to a gapless spectrum in the bulk geometry.
Hence, to verify the bulk--boundary correspondence,
the spectrum in the boundary geometry must also be gapless at this point.
A single solution is insufficient to construct a general solution
compatible with the obc.~\cite{yao1,yokomizo1}
A crossing point between two $b_{i}$ yields two zero-energy solutions,
which should enable us to construct a zero-energy solution
in the boundary geometry under the obc.
We thus expect that $b(\gamma)$ is allowed to cross $b_{i}$
only at such a crossing point.
The second requirement does not uniquely determine $b(\gamma)$,
except at a crossing point,
so that $b(\gamma)$ is arbitrary in a region
between two neighboring gap closing lines in which the two bands have a gap.
Since $\nu(\gamma,b)$ is uniquely determined in such a gapped region,
this arbitrariness does not affect the bulk--boundary correspondence.

The above reasoning implicitly assumes
that a phase transition accompanies a gap closing
(i.e., a touching of the conduction and valence bands) at zero energy,
and that a zero-energy bulk state in the boundary geometry is captured
by the method of a separation of variables.
Here and hereafter, a bulk state represents an eigenstate in the two bands
being distinguishable from chiral edge states.
Although this assumption is plausible, there is an exception.
Let us consider the Chern insulator phase in the boundary geometry,
where the conduction and valence bands are linked by chiral edge states.
If the chiral edge states are destabilized by non-Hermiticity,
the two bands are combined into one band
by a bridge of destabilized chiral edge states (see Fig.~7).
Here, the destabilized states should be regarded as bulk states,
and cannot be captured by the method of a separation of variables.
This is referred to as a peculiar gap closing, which is intrinsic to
two- and three-dimensional non-Hermitian topological systems.
The third requirement is added to take this into account.

When a peculiar gap closing takes place in the boundary geometry,
a chiral edge state at zero energy is transformed into a bulk state
at a transition point to the gapless phase.
Such a state is characterized by
\begin{align}
    \label{eq:b-gamma}
  b = \sqrt{\gamma^{2}+1} + \gamma ,
\end{align}
which is identical to Eq.~(\ref{eq:b-CEs_bg}) at $k = 0$.
Thus, in the nontrivial region, we determine $b(\gamma)$
in the third requirement using Eq.~(\ref{eq:b-gamma}).
At a crossing point of $b(\gamma)$ and a gap closing line,
the bulk geometry becomes gapless.
To verify the bulk--boundary correspondence,
this crossing must coincide with
the destabilization of chiral edge states in the boundary geometry.
This is justified at the end of this section.

We examine the bulk--boundary correspondence
by using the revised recipe.
Possible trajectories of $b(\gamma)$
satisfying the three requirements are shown in Fig.~2.
The trajectories in panels~(a) and (b) are
determined by the third requirement,
whereas the trajectory in panel~(d) is determined by
the first and second requirements.
In panel (c), the trajectory in the region of $\nu = 1$ is determined by
the third requirement,
whereas that in the region of $\nu = 0$ is determined by
the first and second requirements.
For a given $\gamma$, a phase realized in the boundary geometry
is governed by $\nu$ at $b(\gamma)$.
For example, panel~(c) indicates that the phase realized
in the boundary geometry at $M = 2.5$ starts from the trivial phase
($\nu = 0$) at $\gamma = 0$,
changes to the nontrivial phase ($\nu = 1$) with increasing $\gamma$,
and finally enters the gapless phase.
From panels~(a)--(d), we can determine three phase boundaries.
The first one is between the trivial region ($\nu = 0$)
and the nontrivial region ($\nu = 1$),
the second one is between the trivial region ($\nu = 0$)
and the gapless region,
and the third one is between the nontrivial region ($\nu = 1$)
and the gapless region.

%%%%%%%%%%%%%%%%%%
\begin{figure}[btp]
\begin{center}
\includegraphics[height=4.0cm]{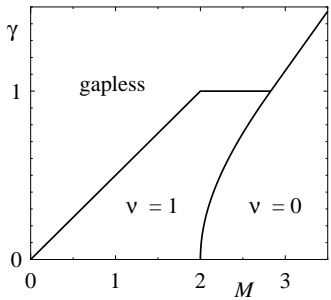}
\end{center}
\caption{
Phase diagram in the boundary geometry, where the regions
designated by $\nu = 0$ and $\nu = 1$ respectively correspond to
the topologically trivial and nontrivial phases,
and outside of them is the gapless phase.
}
\end{figure}
%%%%%%%%%%%%%%%%%%
The phase diagram in the boundary geometry is shown in Fig.~3.
Phase boundaries are determined as follows.
Let us focus on the phase boundary between the trivial and nontrivial regions,
which is at $M = 2$ in the Hermitian limit of $\gamma = 0$.
In the non-Hermitian regime of $\gamma > 0$,
this is determined by the condition of $b_{2}=b_{3}$,
as can be seen from Fig.~2(c).
By solving $b_{2}=b_{3}$, we find
\begin{align}
     \label{eq:PB_tri-nontri}
  M = 2\sqrt{\gamma^{2}+1}
\end{align}
when $2 < M < 2\sqrt{2}$.
This is equivalent to Eq.~(22) of Ref.~\citen{takane1}.
We next consider the phase boundary between the trivial and gapless regions.
As can be seen from Fig.~2(d), this is determined by the condition
of $b_{1} = b_{3}$,
resulting in~\cite{takane1}
\begin{align}
  M = \sqrt{2}\left( \gamma+1 \right)
\end{align}
when $2\sqrt{2} < M$.
We finally consider the phase boundary
between the nontrivial and gapless regions.
The corresponding boundary reported in Ref.~\citen{takane1} is corrected below.
In the range of $M < 2$, this is determined by the condition
of $b(\gamma) = b_{5}$, as can be seen from Fig.~2(a),
resulting in
\begin{align}
    \label{eq:nontrivial-gapless1}
  M = 2\gamma .
\end{align}
The phase boundary in the range of $2 < M < 2\sqrt{2}$
is determined by the condition of $b(\gamma) = b_{0}$,
as can be seen from Fig.~2(c), resulting in
\begin{align}
    \label{eq:nontrivial-gapless2}
  \gamma = 1 .
\end{align}

%%%%%%%%%%%%%%%%%%
\begin{figure}[btp]
\begin{center}
\includegraphics[height=2.0cm]{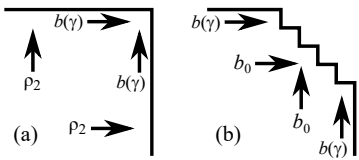}
\end{center}
\caption{
Wavefunction amplitude of chiral edge states near zero energy
increases in the direction specified by each arrow,
where $b(\gamma)$ is the rate of increase per site
in the longitudinal (i.e., propagating or counterpropagating) direction,
and $\rho_{2}$ in panel~(a) and $b_{0}$ in panel~(b)
are the rates of increase per site in the transverse direction. 
}
\end{figure}
%%%%%%%%%%%%%%%%%%
On the boundary specified by Eq.~(\ref{eq:nontrivial-gapless1}),
chiral edge states along the top and right edges [see Fig.~4(a)]
are destabilized.
Penetration of these chiral edge states in the transverse direction
is characterized by $\rho_{2}$ given in Eq.~(\ref{eq:rho-1,2}).
On this boundary, $b(\gamma)$ given in Eq.~(\ref{eq:b-gamma})
becomes equal to $|\rho_{2}|$.
That is, near zero energy, the wavefunction amplitude of the chiral edge states
varies in the longitudinal (i.e., propagating or counterpropagating)
and transverse directions with the same rate of increase.
This is typical of wavefunctions of bulk states.

On the boundary specified by Eq.~(\ref{eq:nontrivial-gapless2}),
chiral edge states along a zigzag edge [see Fig.~4(b)] are destabilized.
Penetration of these chiral edge states in the transverse direction
is characterized by $b_{0}$ given in Eq.~(\ref{eq:b_0}).
On this boundary, $b(\gamma)$ given in Eq.~(\ref{eq:b-gamma})
becomes equal to $b_{0}$,
indicating that the wavefunction amplitude of the chiral edge states varies
in the longitudinal and transverse directions with the same rate of increase.
Again, this is typical of wavefunctions of bulk states.
This behavior is visible in the boundary geometry
when its edge contains a zigzag structure (data not shown).
If a zigzag structure is absent,
as in the square-shaped system considered in Sect.~6,
the destabilization of chiral edge states is difficult to detect
in the behavior of a wavefunction.
Despite this, we can detect the destabilization in the square-shaped system
by observing complex spectra [see Fig.~6(d)].

\section{Numerical Results}

We examine the phase boundaries shown in Fig.~3
by comparing them with spectra in the boundary geometry
of $160 \times 160$ sites under the obc.
We determine the spectrum of $H$ for a given set of parameters
by computing eigenvalues of $\tilde{H} = \Lambda^{-1}H\Lambda$,
where $\Lambda$ is the representative matrix of a similarity transformation:
\begin{align}
   \Lambda
   = \sum_{m,n} |m,n \rangle b^{m+n} \langle m,n| .
\end{align}
The accuracy of computation is improved
if we appropriately set the value of $b$.
Numerical libraries of LAPACK and z-Pares~\cite{sakurai,futamura}
with MUMPS~\cite{amestoy} are used in the computation.

%%%%%%%%%%%%%%%%%%
\begin{figure}[btp]
\begin{tabular}{cc}
\begin{minipage}{0.5\hsize}
\begin{center}
\hspace{-10mm}
\includegraphics[height=2.4cm]{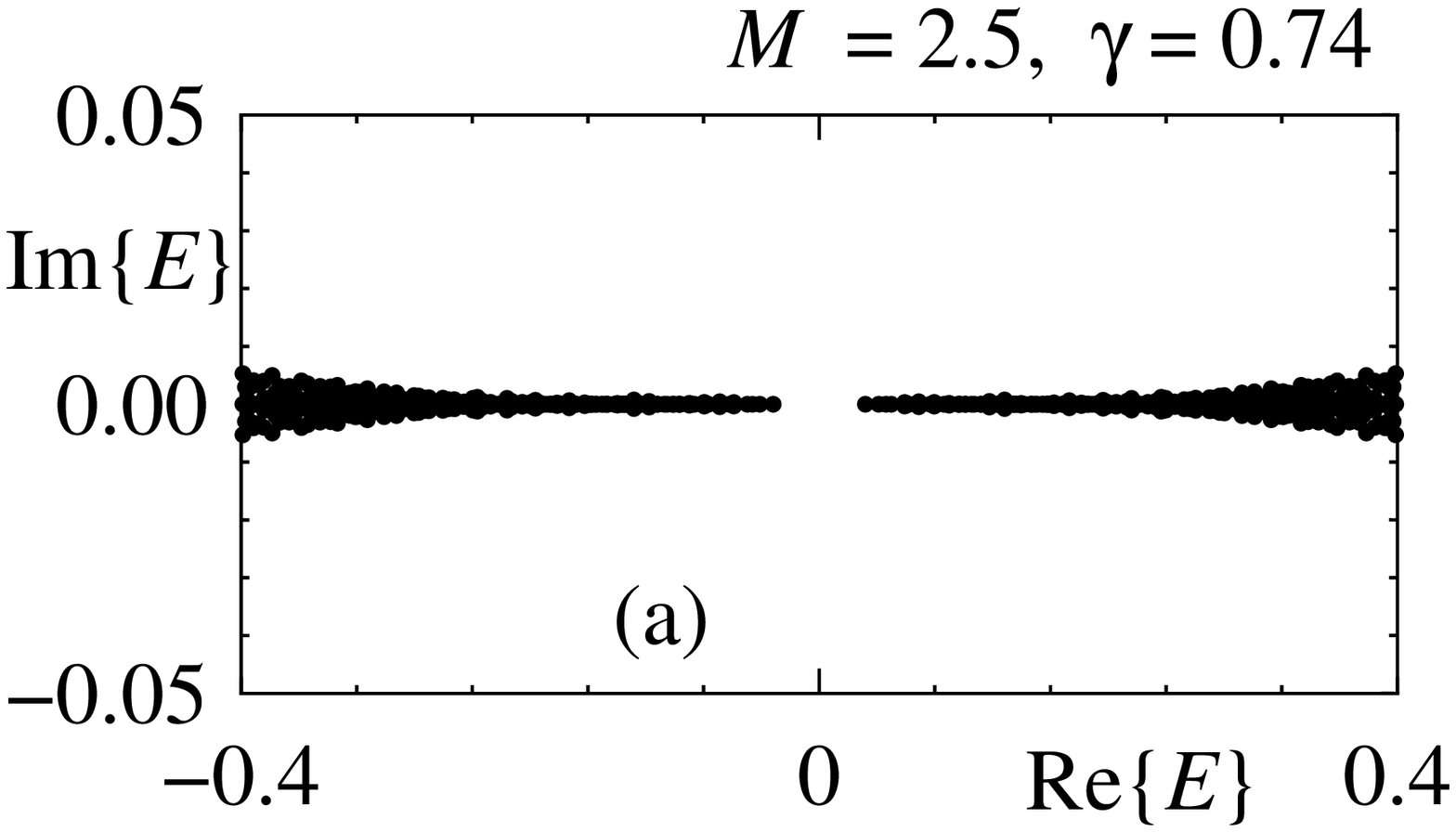}
\end{center}
\end{minipage}
\begin{minipage}{0.5\hsize}
\begin{center}
\hspace{-5mm}
\includegraphics[height=2.4cm]{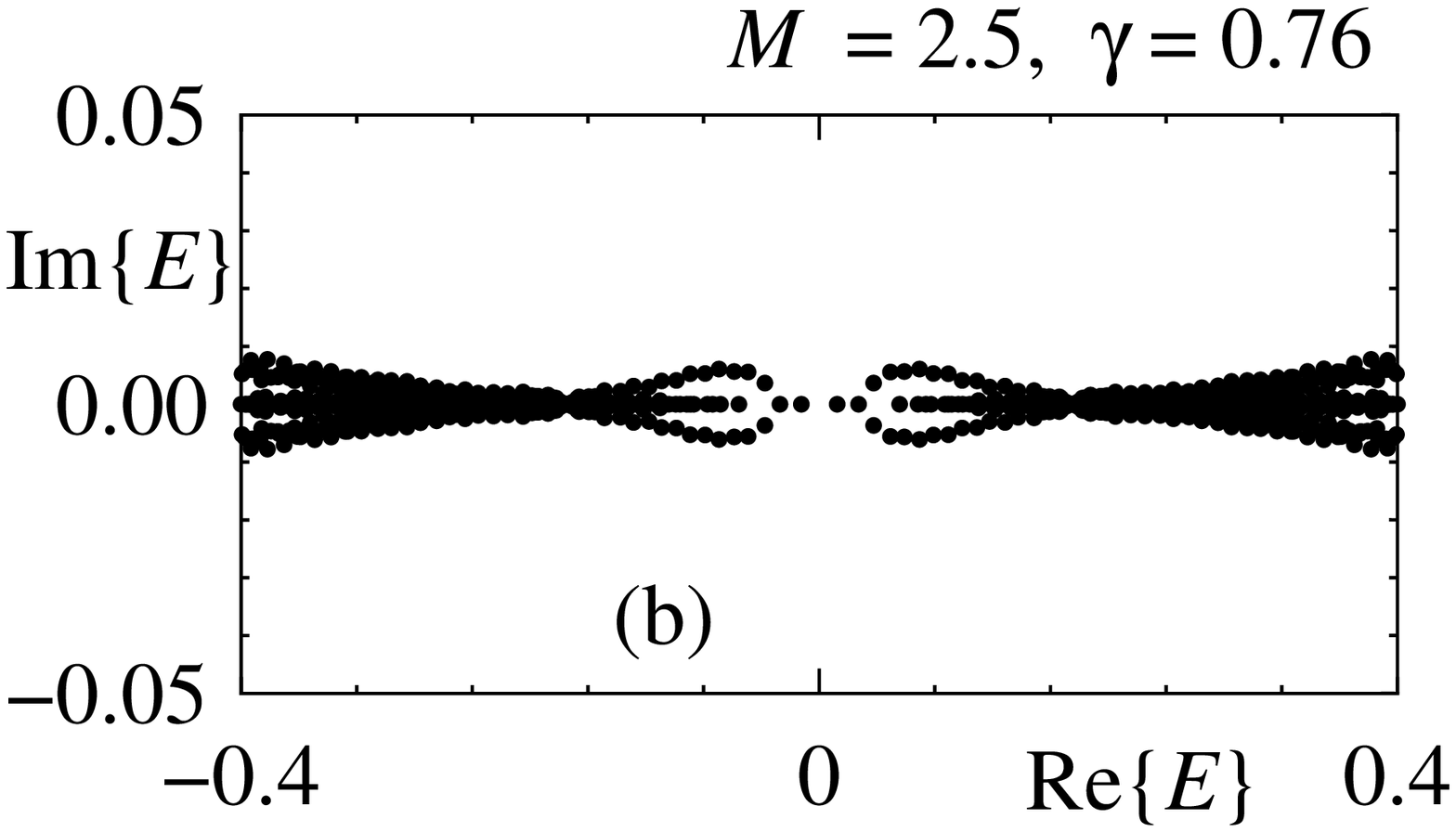}
\end{center}
\end{minipage}
\end{tabular}
\begin{tabular}{cc}
\begin{minipage}{0.5\hsize}
\begin{center}
\hspace{-10mm}
\includegraphics[height=2.4cm]{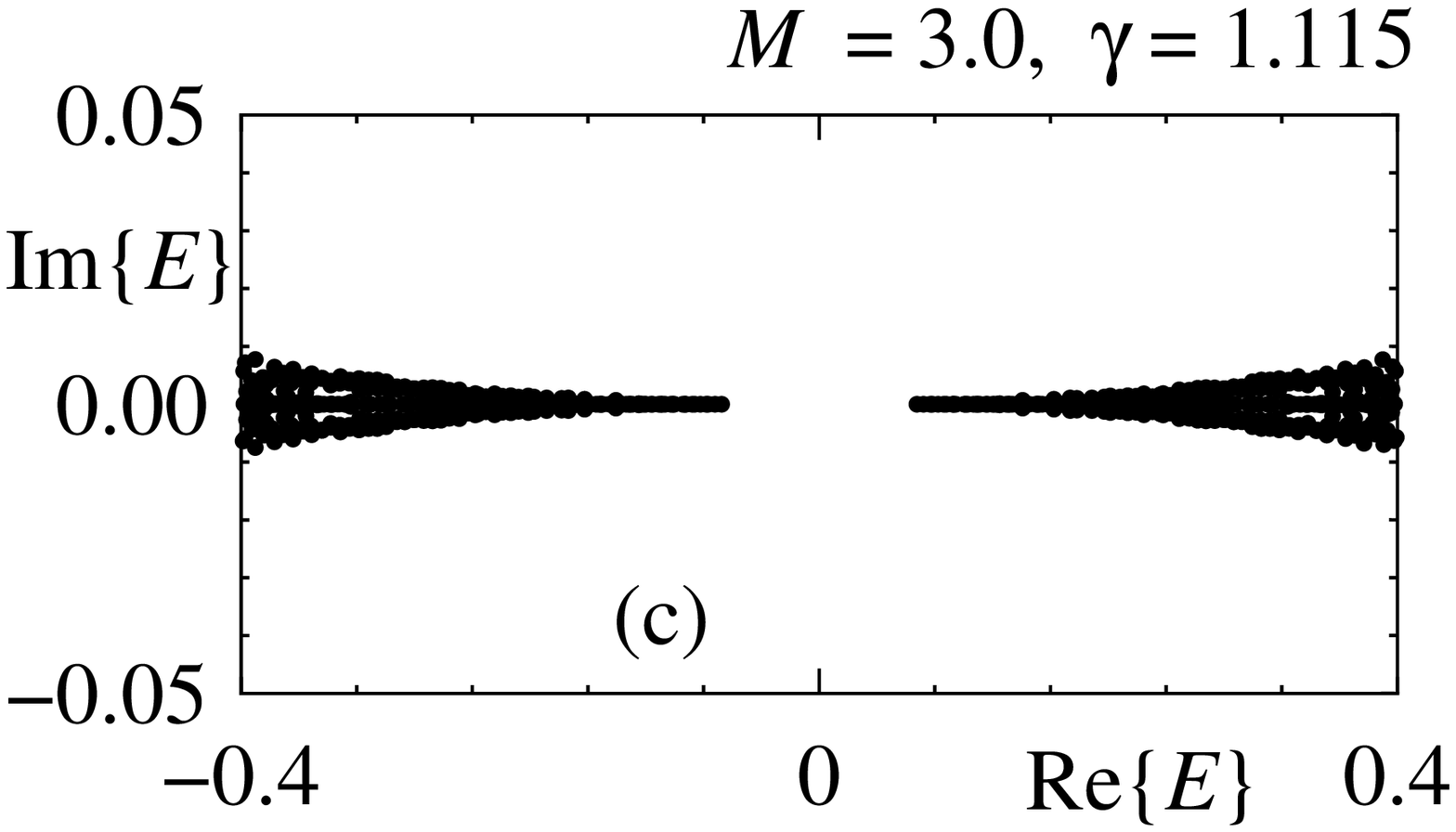}
\end{center}
\end{minipage}
\begin{minipage}{0.5\hsize}
\begin{center}
\hspace{-5mm}
\includegraphics[height=2.4cm]{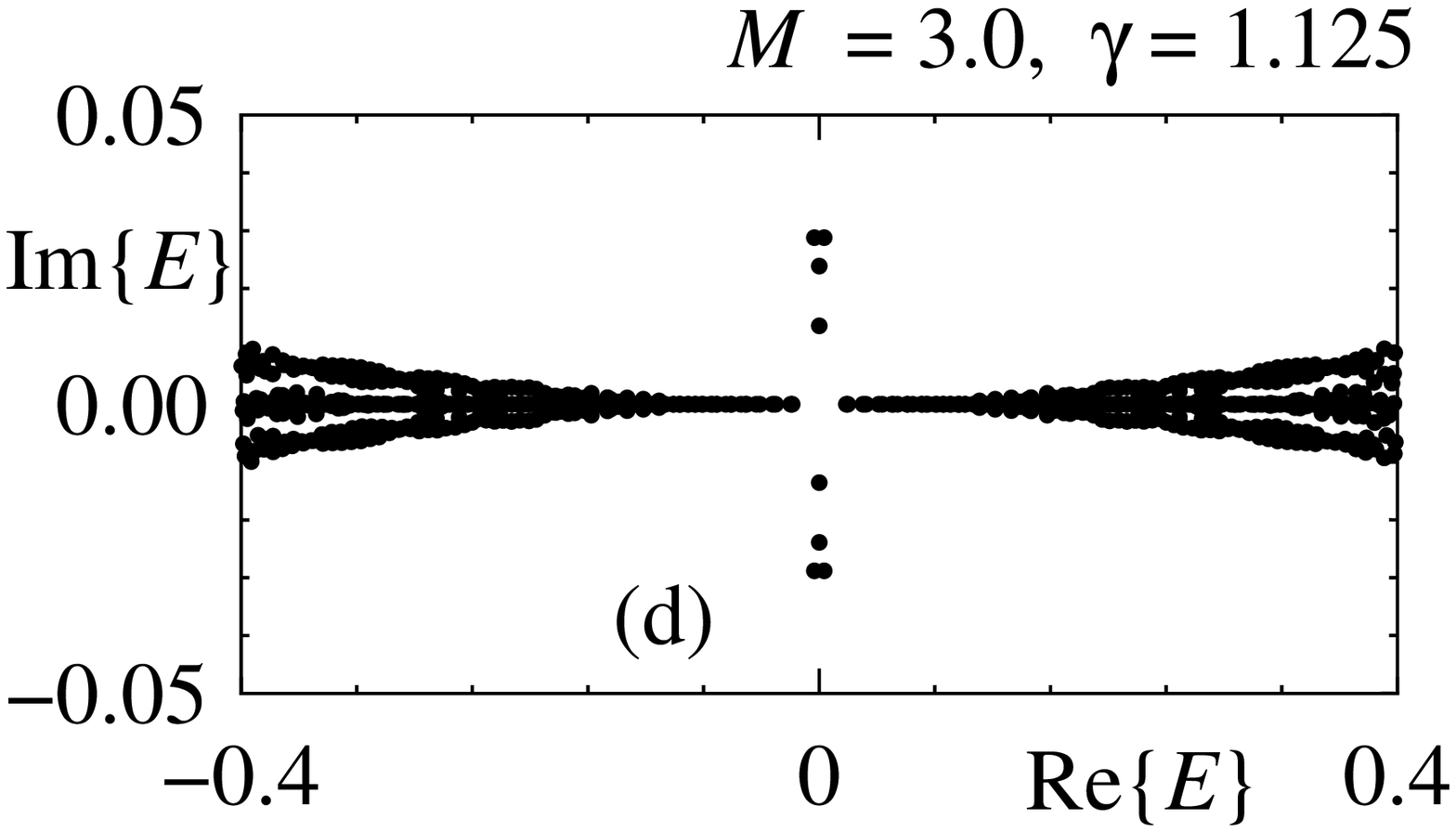}
\end{center}
\end{minipage}
\end{tabular}
\caption{
Spectra in the boundary geometry of $160 \times 160$ sites
at $M = 2.5$
with $\gamma =$ (a) $0.74$ and (b) $0.76$,
where the phase boundary is at $\gamma_{\rm c} = 0.75$,
and those at $M = 3.0$
with $\gamma =$ (c) $1.115$ and (d) $1.125$,
where the phase boundary is at $\gamma_{\rm c} \approx 1.121$.
}
\end{figure}
%%%%%%%%%%%%%%%%%%
Let us briefly consider
the phase boundary between the trivial and nontrivial regions
and that between the trivial and gapless regions.
Figure~5 shows the spectra in the cases of $M = 2.5$ and $3.0$.
In the case of $M = 2.5$, the phase boundary between
the trivial and nontrivial regions is at $\gamma_{\rm c} = 0.75$.
The spectrum at $\gamma = 0.74$ [panel (a)] has
a small gap between the conduction and valence bands,
whereas the spectrum at $\gamma = 0.76$ [panel (b)] contains several dots
between the two bands, which represent chiral edge states.
These results are consistent with $\gamma_{\rm c} = 0.75$ and
indicate that the system changes from the trivial phase
to the nontrivial phase at $\gamma = \gamma_{\rm c}$.
In the case of $M = 3.0$, the phase boundary between
the trivial and gapless regions is at $\gamma_{\rm c} \approx 1.121$.
The spectrum at $\gamma = 1.115$ [panel (c)] has a gap between the two bands,
whereas the spectrum at $\gamma = 1.125$ [panel (d)] becomes gapless
with several states appearing on the imaginary axis.
These results are consistent with $\gamma_{\rm c} \approx 1.121$ and
indicate that the system changes from the trivial phase
to the gapless phase at $\gamma = \gamma_{\rm c}$.

%%%%%%%%%%%%%%%%%%
\begin{figure}[btp]
\begin{tabular}{cc}
\begin{minipage}{0.5\hsize}
\begin{center}
\hspace{-10mm}
\includegraphics[height=2.4cm]{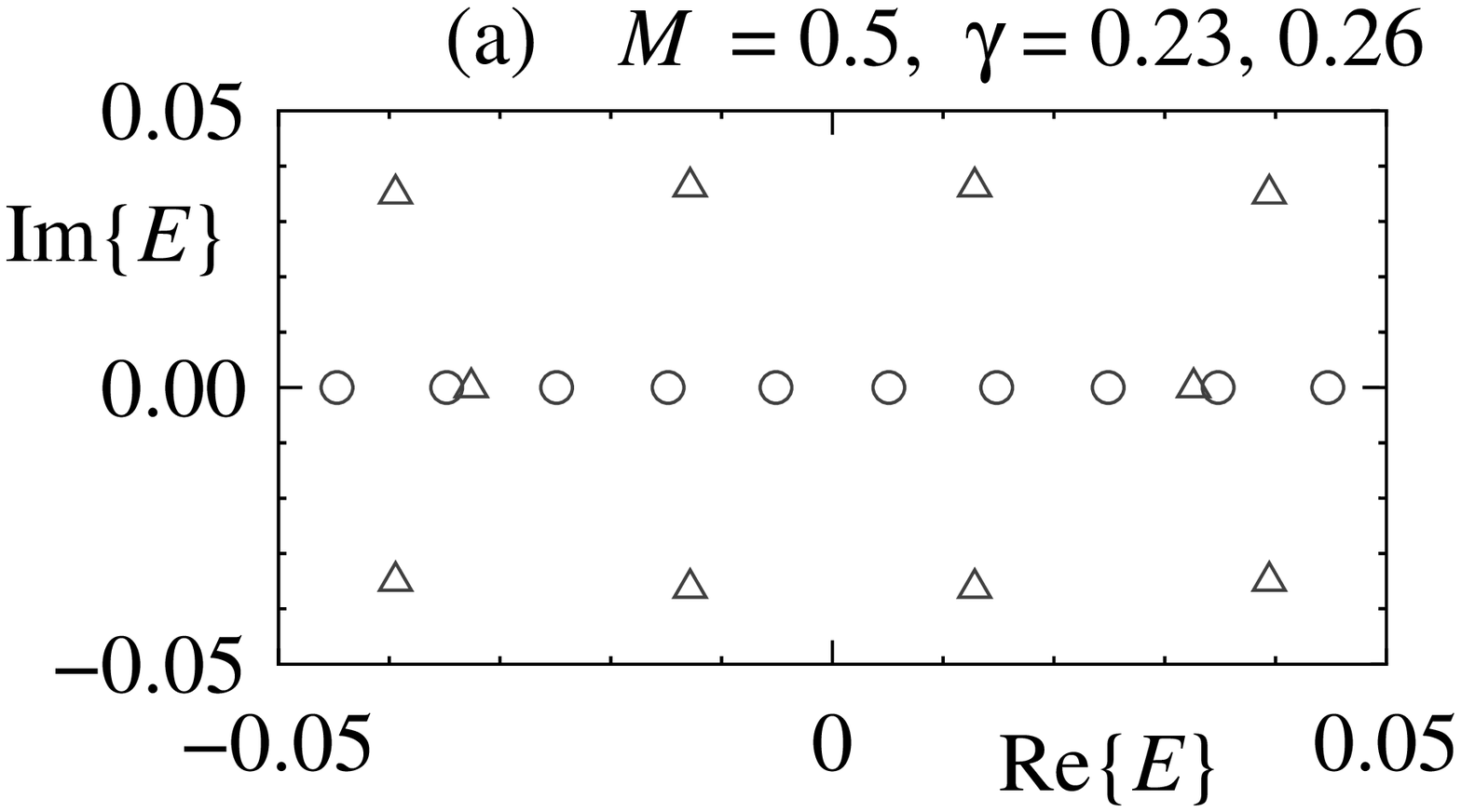}
\end{center}
\end{minipage}
\begin{minipage}{0.5\hsize}
\begin{center}
\hspace{-5mm}
\includegraphics[height=2.4cm]{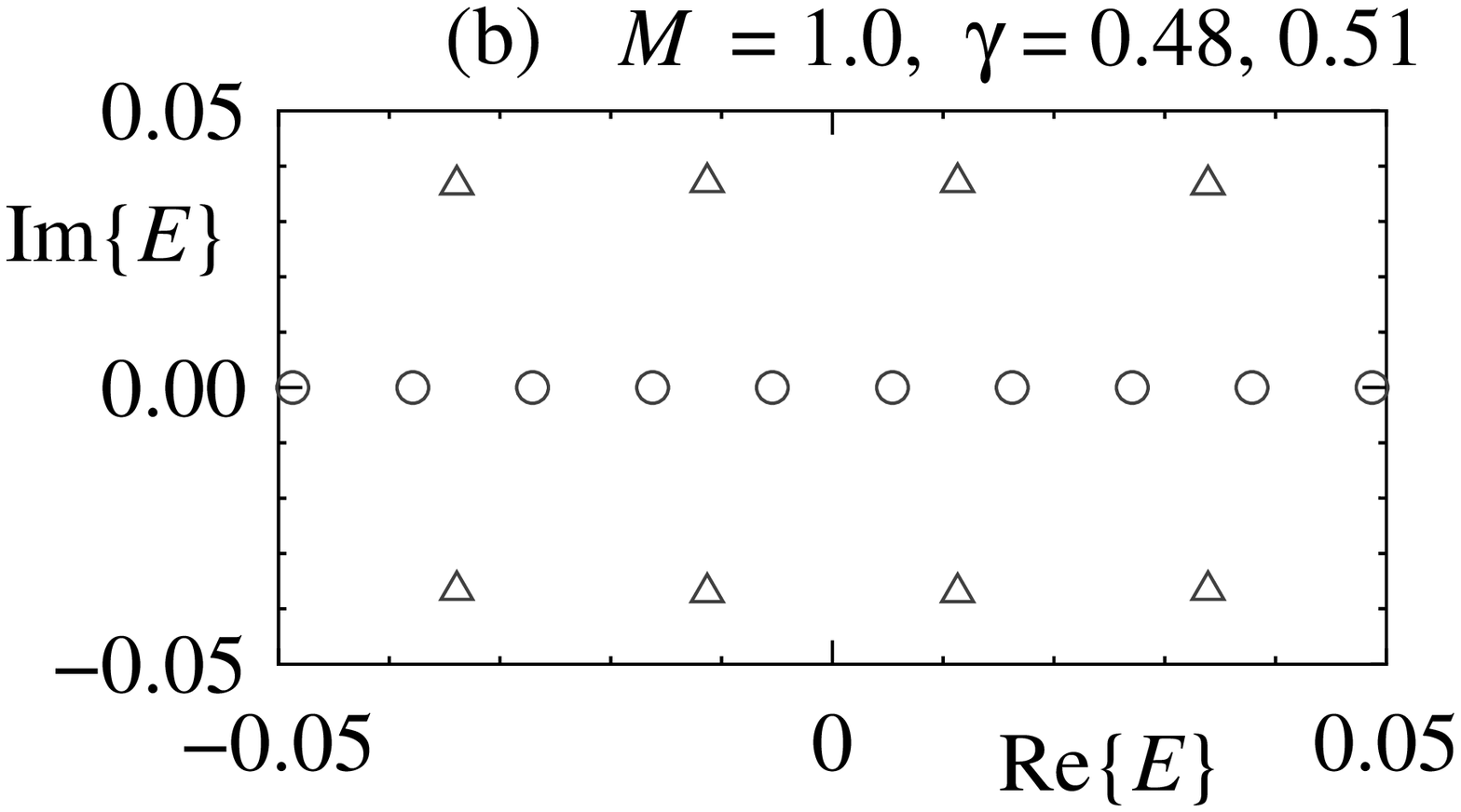}
\end{center}
\end{minipage}
\end{tabular}
\begin{tabular}{cc}
\begin{minipage}{0.5\hsize}
\begin{center}
\hspace{-10mm}
\includegraphics[height=2.4cm]{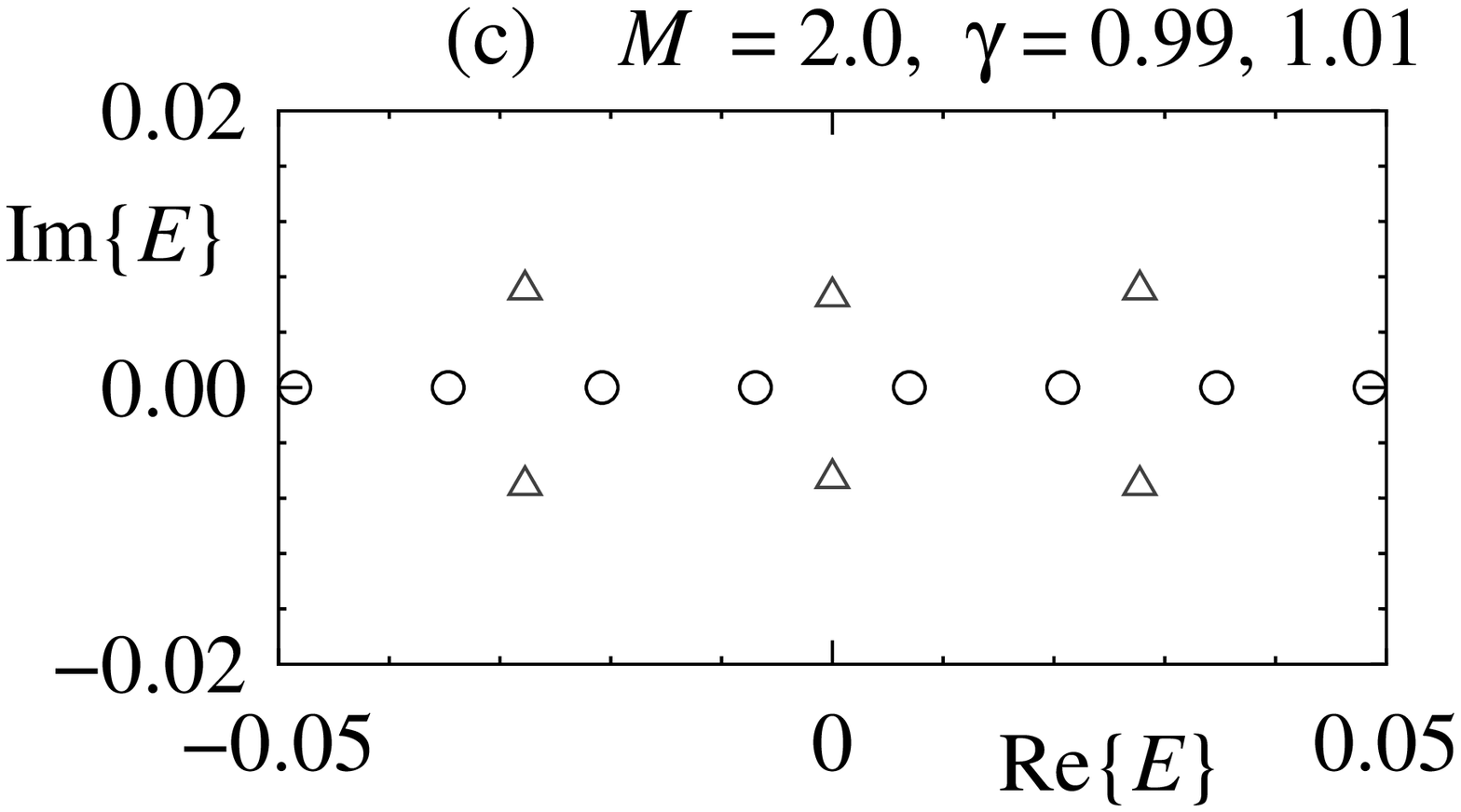}
\end{center}
\end{minipage}
\begin{minipage}{0.5\hsize}
\begin{center}
\hspace{-5mm}
\includegraphics[height=2.4cm]{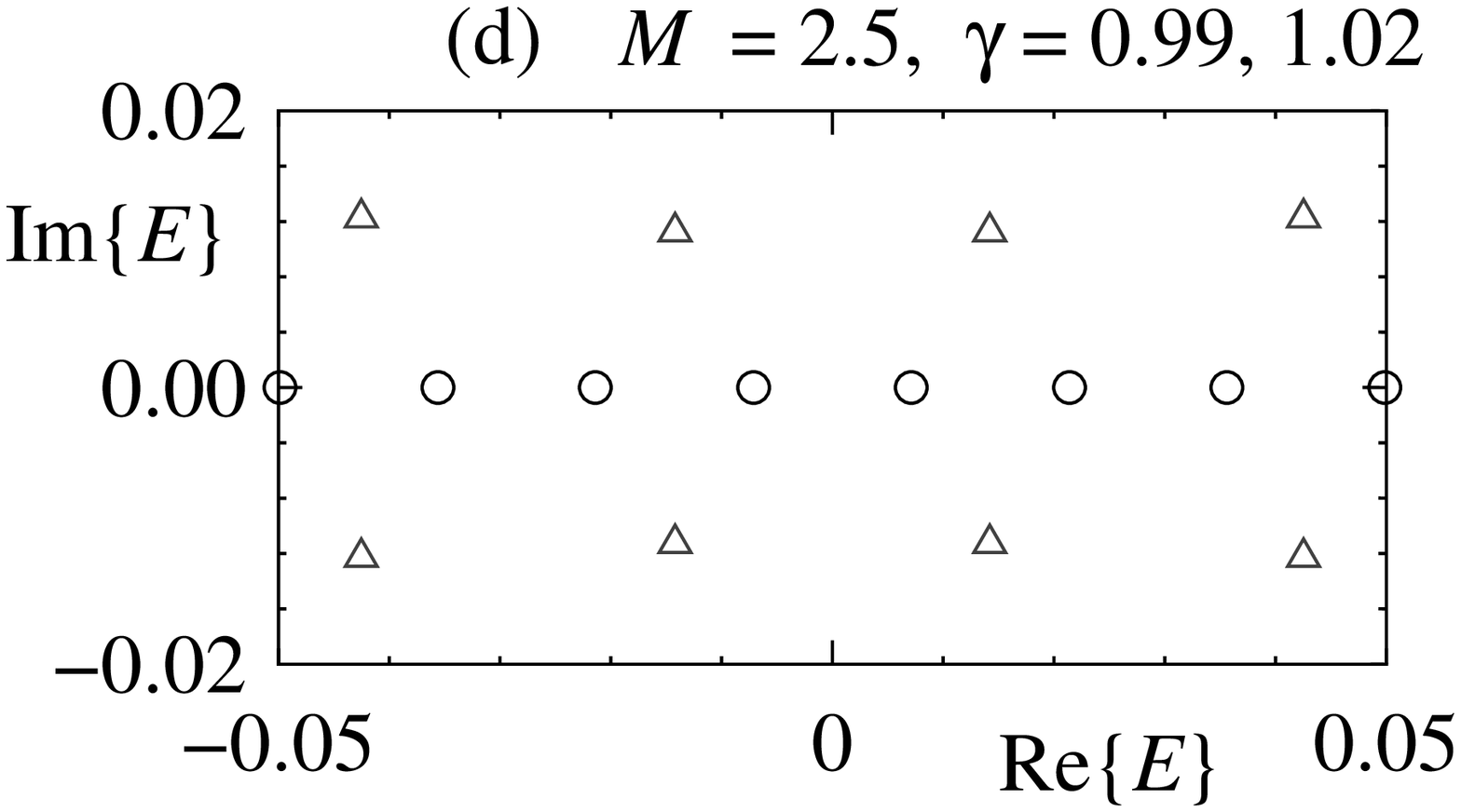}
\end{center}
\end{minipage}
\end{tabular}
\caption{
Spectra in the boundary geometry of $160 \times 160$ sites
(a) at $M = 0.5$ with $\gamma = 0.23$ (circles) and $0.26$ (triangles),
where the phase boundary is at $\gamma_{\rm c} = 0.25$,
(b) at $M = 1.0$ with $\gamma = 0.48$ (circles) and $0.51$ (triangles),
where the phase boundary is at $\gamma_{\rm c} = 0.50$,
(c) at $M = 2.0$ with $\gamma = 0.99$ (circles) and $1.01$ (triangles),
where the phase boundary is at $\gamma_{\rm c} = 1.00$, and
(d) at $M = 2.5$ with $\gamma = 0.99$ (circles) and $1.02$ (triangles),
where the phase boundary is at $\gamma_{\rm c} = 1.00$.
}
\end{figure}
%%%%%%%%%%%%%%%%%%
We hereafter examine the phase boundary
between the nontrivial and gapless regions.
Figure~6 shows the spectra in the cases of $M = 0.5$, $1.0$, $2.0$, and $2.5$.
Only the low energy region of $-0.05 \le {\rm Re}\{E\} \le 0.05$
near the real axis is shown so that we focus on chiral edge states.
As discussed in Sect.~4, chiral edge states are stable
only when their energies are real.
In the case of $M = 0.5$, the phase boundary between
the nontrivial and gapless regions is at $\gamma_{\rm c} = 0.25$.
In panel (a) with $M = 0.5$, eigenvalues of energy at $\gamma = 0.23$ are
almost equally distributed on the real axis,
whereas those at $\gamma = 0.26$ deviate from the real axis,
reflecting the destabilization of chiral edge states.
These results are consistent with $\gamma_{\rm c} = 0.25$ and
indicate that the system changes from the nontrivial phase
to the gapless phase at $\gamma = \gamma_{\rm c}$.
Moreover, in panels~(b), (c), and (d), we observe behavior similar to this;
eigenvalues corresponding to chiral edge states are almost equally
distributed on the real axis when $\gamma < \gamma_{\rm c}$ and deviate
from the real axis once $\gamma$ exceeds $\gamma_{\rm c}$.
These results justify the phase boundary
between the nontrivial and gapless regions determined in Sect.~5.

Finally, we show typical spectra in the boundary geometry
to give an example of the peculiar gap closing pointed out in Sect.~5.
Figure~7 shows spectra at $M = 2.5$ with (a) $\gamma = 0.95$
in the nontrivial phase and (b) $\gamma = 1.05$ in the gapless phase.
In the nontrivial phase, we consider that
the conduction and valence bands are separated by a gap
although they are linked by chiral edge states.
This is because we should distinguish the chiral edge states
from bulk states, as in the Hermitian limit.
In the gapless phase, it should be recognized that the two bands are
combined into one band by a bridge of destabilized chiral edge states.
This is because the destabilized chiral edge states are bulk states.
%%%%%%%%%%%%%%%%%%
\begin{figure}[btp]
\begin{tabular}{cc}
\begin{minipage}{0.5\hsize}
\begin{center}
\hspace{-10mm}
\includegraphics[height=3.5cm]{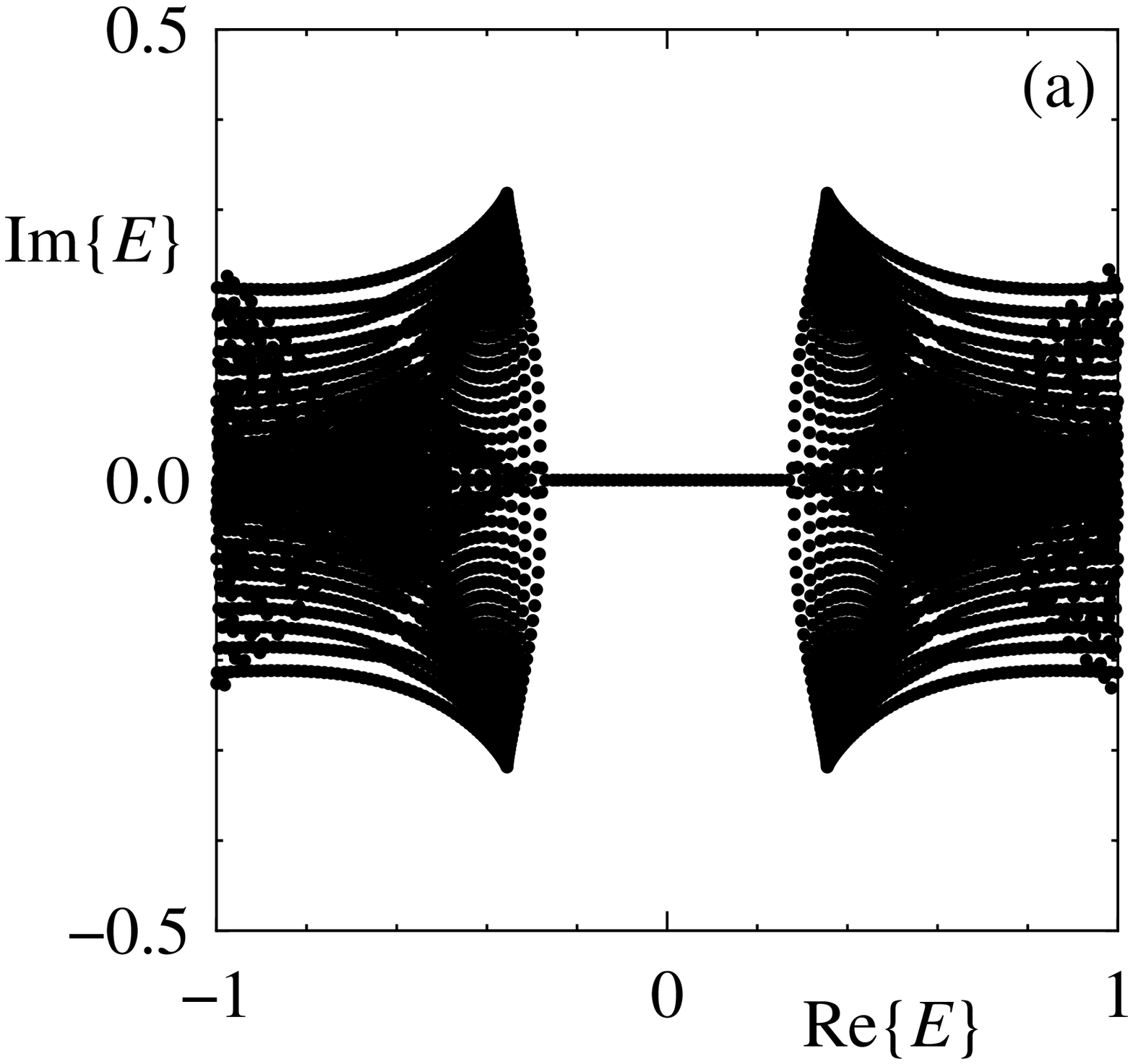}
\end{center}
\end{minipage}
\begin{minipage}{0.5\hsize}
\begin{center}
\hspace{-5mm}
\includegraphics[height=3.5cm]{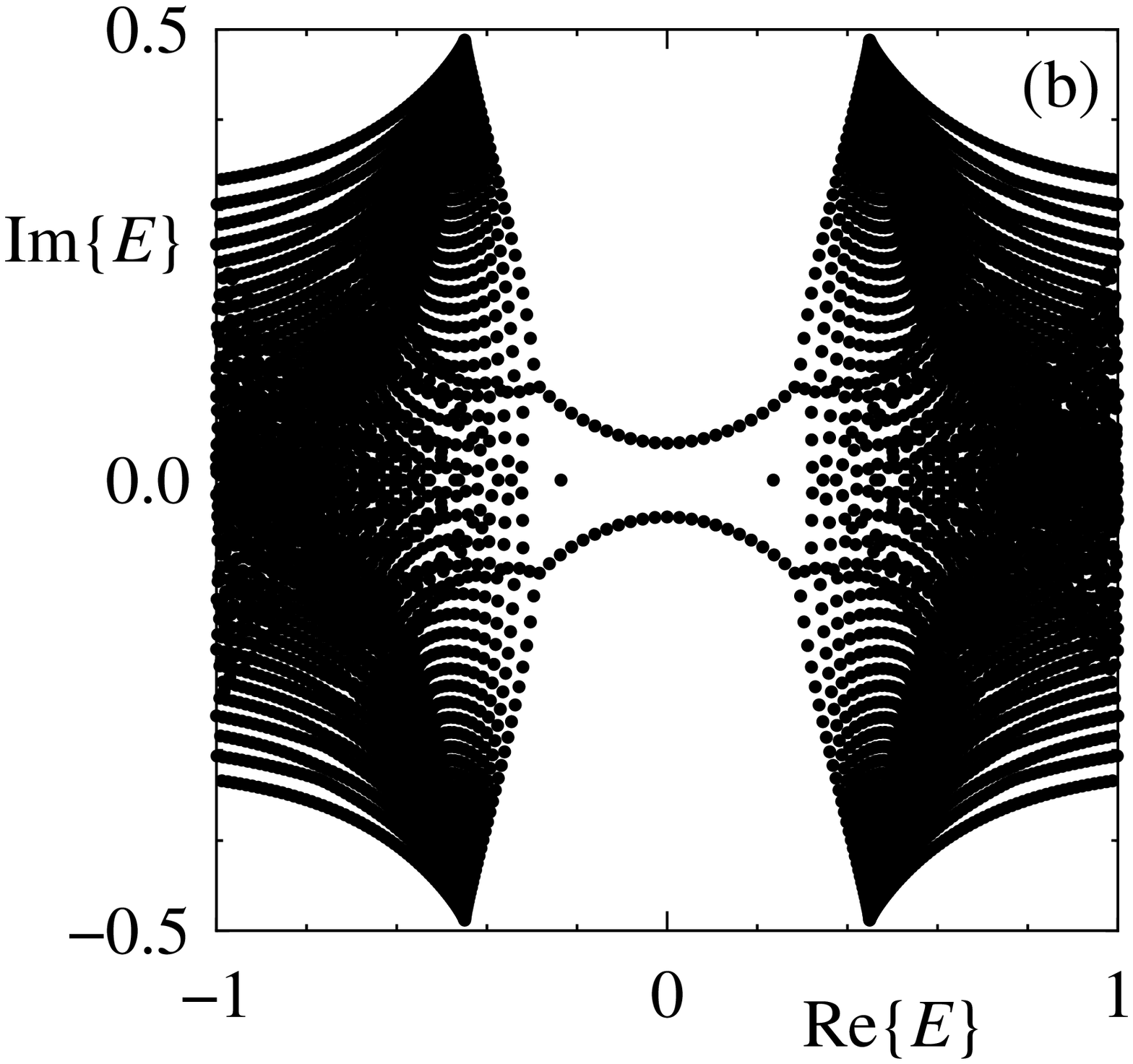}
\end{center}
\end{minipage}
\end{tabular}
\caption{
Boundary spectra in the case of $M = 2.5$
with $\gamma =$ (a) $0.95$ and (b) $1.05$.
Panel~(a) shows the spectrum in the nontrivial phase
with chiral edge states on the real axis.
Panel~(b) shows the spectrum in the gapless phase,
where chiral edge states are destabilized and split
into two branches off the real axis.
Since destabilized chiral edge states are regarded as bulk states,
we should consider that the conduction and valence bands
are combined into one band
by a bridge of destabilized chiral edge states in the gapless phase.
}
\end{figure}
%%%%%%%%%%%%%%%%%%

\section{Summary and Discussion}

In two- and three-dimensional non-Hermitian topological systems,
topological boundary states can be destabilized by non-Hermiticity
and transformed into bulk states that bridge conduction and valence bands,
resulting in a peculiar gap closing.
We revised the recipe of
non-Hermitian bulk--boundary correspondence~\cite{takane1}
to take this into account,
and applied it to the model of a non-Hermitian Chern insulator.~\cite{yao2}

In the bulk--boundary correspondence demonstrated in Sect.~5,
the argument on the bulk geometry uses Eq.~(\ref{eq:b-gamma}) as input.
This additional input is necessary to take account of the peculiar gap closing
due to the destabilization of chiral edge states.
We showed that, with this input, the bulk--boundary correspondence
correctly describes the boundary property specified in Fig.~3
including the destabilization of chiral edge states.
Note that it is very difficult to derive the boundary property
from the eigenvalue equation in the boundary geometry.
This indicates that the bulk--boundary correspondence is useful
in considering the boundary property of non-Hermitian topological systems.

Finally, we consider the applicability of the recipe described in Sect.~5
to a general non-Hermitian topological system in two dimensions.
A phase transition of the system in the boundary geometry
accompanies a gap closing, which denotes a direct touching of
conduction and valence bands in a narrow sense and includes
a peculiar gap closing
(i.e., destabilization of boundary states that link the two bands)
in a wide sense.
The applicability crucially depends on low-energy states
in the boundary geometry, in particular,
a bulk state that is located at a touching point between the two bands.
In the case of a peculiar gap closing, a bulk state
should be replaced with a destabilized boundary state.
This recipe is applicable
if the bulk state at a touching point can be captured by
the method of a separation of variables
or if a non-Hermitian skin effect on it can be characterized
in a quantitative manner as in Eq.~(\ref{eq:b-gamma}).

\section*{Acknowledgment}

This work was supported by JSPS KAKENHI Grant Number JP21K03405.

\section*{Appendix}

We show that the dispersion relations given in Eqs.~(\ref{eq:disp-b}),
(\ref{eq:disp-u}), (\ref{eq:disp-l}), and (\ref{eq:disp-r}) are valid
even when $|\rho_{1}| < 1 < |\rho_{2}|$ does not hold.
In our system, the condition of $1 < |\rho_{2}|$ widely holds
but the condition of $|\rho_{1}| < 1$ breaks down with increasing $\gamma$.
When $1 < |\rho_{1}|$, we need to consider the boundary conditions at $n = 0$
and $n = N+1$ on equal footing.
If this is done, the eigenvalue of energy $E_{\bot}$ deviates from $0$.
Let us denote this deviation as $\delta E$
and assume that $\delta E$ is sufficiently small.
We obtain the dispersion relations of chiral edge states
by using the method given in Ref.~\citen{okamoto}.
Up to the first order in $\delta E$, the eigenvectors are modified as
\begin{align}
  |\tilde{\psi}_{1}^{R}\rangle
  & = \frac{1}{\sqrt{2}}
                    \left[ \begin{array}{c}
                              1 \\
                              1
                     \end{array}
                    \right]
    + \xi_{1}
      \frac{1}{\sqrt{2}}
                    \left[ \begin{array}{c}
                              1 \\
                              -1
                     \end{array}
                    \right] ,
      \\
  |\tilde{\psi}_{2}^{R}\rangle
  & = \frac{1}{\sqrt{2}}
                    \left[ \begin{array}{c}
                              1 \\
                              -1
                     \end{array}
                    \right]
    + \xi_{2}
      \frac{1}{\sqrt{2}}
                    \left[ \begin{array}{c}
                              1 \\
                              1
                     \end{array}
                    \right] ,
\end{align}
with
\begin{align}
  \xi_{j}
  = \frac{\delta E}
         {2\left( \tilde{M}-\frac{\rho_{j}+\rho_{j}^{-1}}{2} \right)} ,
\end{align}
where $j = 1$, $2$.
We can ignore corrections to $\rho_{1}$ and $\rho_{2}$,
which are on the order of $\delta E^{2}$.
Taking these into account, we construct right eigenfunctions of
chiral edge states by superposing the following fundamental solutions:
\begin{align}
    |\tilde{\Psi}_{1}^{R}\rangle
 & = \sum_{m}(be^{ik_{x}a})^{m}\sum_{n=0}^{N+1}
     \rho_{1}^{n}|m,n \rangle \cdot |\tilde{\psi}_{1}^{R}\rangle ,
        \\
    |\tilde{\Psi}_{2}^{R}\rangle
 & = \sum_{m}(be^{ik_{x}a})^{m}\sum_{n=0}^{N+1}
     \rho_{2}^{n}|m,n \rangle \cdot |\tilde{\psi}_{2}^{R}\rangle ,
        \\
    |\Psi_{\rm bottom}^{R}\rangle
 & = \sum_{m}(be^{ik_{x}a})^{m}|m,0 \rangle \cdot |\psi_{1}^{R}\rangle ,
     \\
    |\Psi_{\rm top}^{R}\rangle
 & = \sum_{m}(be^{ik_{x}a})^{m}|m,N+1 \rangle \cdot |\psi_{2}^{R}\rangle .
\end{align}
As a result, we find two eigenfunctions of
$|\Psi_{+}^{R}\rangle$ with $E = \delta E$ and
$|\Psi_{-}^{R}\rangle$ with $E = -\delta E$ under the condition of
\begin{align}
     \label{eq:condition-chiral_edge}
   \xi_{1}\xi_{2} = \left(\frac{\rho_{1}}{\rho_{2}}\right)^{N+1} .
\end{align}
These eigenfunctions are given by
\begin{align}
    |\Psi_{\pm}^{R}\rangle
  & = c \sum_{m}(be^{ik_{x}a})^{m}
        \nonumber \\
  & \hspace{-8mm}
      \times \sum_{n=1}^{N}|m,n \rangle\cdot
      \Biggl[ \frac{1}{\sqrt{2}}
             \left[ \begin{array}{c}
                            1 \\
                            1
                    \end{array}
             \right]
             \rho_{1}^{N+1}\left( \frac{\rho_{1}^{n}}{\rho_{1}^{N+1}}
                                 -\frac{\rho_{2}^{n}}{\rho_{2}^{N+1}}
                           \right)
        \nonumber \\
  & \hspace{13mm} 
         \pm \xi_{1}
             \frac{1}{\sqrt{2}}
             \left[ \begin{array}{c}
                            1 \\
                            -1
                    \end{array}
             \right]
             \left( \rho_{1}^{n}-\rho_{2}^{n} \right)
      \Biggr] ,
\end{align}
where $c$ is a normalization constant.
The left eigenfunctions corresponding to $|\Psi_{\pm}^{R}\rangle$ are given by
\begin{align}
    \langle \Psi_{\pm}^{L}|
  & = c \sum_{m}(be^{ik_{x}a})^{-m}
        \nonumber \\
  & \hspace{-8mm}
      \times \sum_{n=1}^{N}
      \Biggl[ 
             \left( \frac{\rho_{1}^{N+1}}{\rho_{1}^{n}}
                   -\frac{\rho_{2}^{N+1}}{\rho_{2}^{n}}
             \right)
             \frac{1}{\sqrt{2}}^{t}\!
             \left[ \begin{array}{c}
                            1 \\
                            1
                    \end{array}
             \right]
        \nonumber \\
  & \hspace{-1mm} 
         \pm \xi_{2}\rho_{2}^{N+1}
             \left( \frac{1}{\rho_{1}^{n}}
                    -\frac{1}{\rho_{2}^{n}} \right)
             \frac{1}{\sqrt{2}}^{t}\!
             \left[ \begin{array}{c}
                            1 \\
                            -1
                    \end{array}
             \right]
      \Biggr]
      \cdot \langle m,n| .
\end{align}
The normalization constant is determined as
\begin{align}
   c = \frac{1}
            {\sqrt{2N\rho_{1}^{N+1}
                   \left(2N + \frac{\rho_{1}^{N+1}+\rho_{2}^{N+1}}
                                   {\rho_{1}-\rho_{2}}
                              \left(\frac{1}{\rho_{1}^{N}}
                                    - \frac{1}{\rho_{2}^{N}}\right)
                   \right)}}
\end{align}
so that $\langle \Psi_{\pm}^{L}|\Psi_{\pm}^{R}\rangle = 1$.
In the space spanned by $|\Psi_{\pm}^{R}\rangle$ and $\langle \Psi_{\pm}^{L}|$,
the Hamiltonian $H$ is reduced to
\begin{align}
  \left[ \begin{array}{cc}
           \langle \Psi_{+}^{L}|H|\Psi_{+}^{R}\rangle
           & \langle \Psi_{+}^{L}|H|\Psi_{-}^{R}\rangle \\
           \langle \Psi_{-}^{L}|H|\Psi_{+}^{R}\rangle
           & \langle \Psi_{-}^{L}|H|\Psi_{-}^{R}\rangle
         \end{array}
  \right]
  =
  \left[ \begin{array}{cc}
           \delta E & \eta_{x} \\
           \eta_{x} & -\delta E
         \end{array}
  \right] .
\end{align}
The eigenvalues of energy are obtained as
\begin{align}
       \label{eq:disp-overlap}
  E = \pm \sqrt{\eta_{x}^{2}+\delta E^{2}} .
\end{align}
Equation~(\ref{eq:condition-chiral_edge}) yields
\begin{align}
  \delta E^{2}
  = 4\left( \tilde{M}-\frac{\rho_{1}+\rho_{1}^{-1}}{2} \right)
      \left( \tilde{M}-\frac{\rho_{2}+\rho_{2}^{-1}}{2} \right)
      \left( \frac{\rho_{1}}{\rho_{2}} \right)^{N+1} ,
\end{align}
which shows that $\delta E^{2}$ vanishes as long as
$N$ is sufficiently large and $|\rho_{1}| < |\rho_{2}|$.
When $\delta E^{2}$ is ignorable, Eq.~(\ref{eq:disp-overlap}) is
reduced to Eqs.~(\ref{eq:disp-b}) and (\ref{eq:disp-u}).

\end{document}